\documentclass[journal=jacsat,manuscript=article]{achemso}

\usepackage[version=4]{mhchem}
\usepackage{siunitx}
\usepackage{amsmath,amssymb}
\usepackage{graphicx}
\usepackage{booktabs,makecell}
\usepackage{placeins}
\usepackage{adjustbox}
\usepackage{multirow}
\usepackage{xr-hyper}
\usepackage{hyperref}
\usepackage{xcolor}

\newcommand{\RomanNumeralCaps}[1]
    {\MakeUppercase{\romannumeral #1}}

\title{Mapping the Optical Landscape of a Squaraine Molecule in the Visible and Ultraviolet Energy Range}

\author{Narges Taghizade}
\affiliation{Institute of Physics, University of Graz, Graz 8010, Austria}

\author{Robert Schwarzl}
\affiliation{Institute of Experimental Physics, Graz University of Technology, Graz 8010, Austria}

\author{Frederik Leinenbach}
\affiliation{Institute of Experimental Physics, Graz University of Technology, Graz 8010, Austria}

\author{Maximilian Jeindl}
\affiliation{Department of Applied Physics and Science Education, Eindhoven University of Technology, Eindhoven 5612 AP, The Netherlands}

\author{Marvin F. Schumacher}
\affiliation{Kekulé-Institute for Organic Chemistry and Biochemistry, University of Bonn, Bonn 53121, Germany}

\author{Brunella Bardi}
\affiliation{Department of Chemistry, Life Sciences and Environmental Sustainability, University of Parma, 43124 Parma, Italy}

\author{Arne L{\"u}tzen}
\affiliation{Kekulé-Institute for Organic Chemistry and Biochemistry, University of Bonn, Bonn 53121, Germany}

\author{Manuela Schiek}
\affiliation{Institute of Physical Chemistry and Linz Institute for Organic Solar Cells \& Center for Surface and Nano-Analytics, Johannes Kepler University Linz, Linz 4040, Austria}

\author{Andreas W. Hauser}
\affiliation{Institute of Experimental Physics, Graz University of Technology, Graz 8010, Austria}

\author{Peter Puschnig}
\affiliation{Institute of Physics, University of Graz, Graz 8010, Austria}
\email{peter.puschnig@uni-graz.at}

\author{Markus Koch}
\affiliation{Institute of Experimental Physics, Graz University of Technology, Graz 8010, Austria}
\email{markus.koch@tugraz.at}

\author{Andreas Windischbacher}
\affiliation{Institute of Physics, University of Graz, Graz 8010, Austria}
\email{andreas.windischbacher@uni-graz.at}

\begin{document}

\begin{abstract}
Although squaraine dyes are commonly praised as candidates for light-based applications, little is known about their excited state landscape beyond the low-energy visible light region.
Our work aims for an improved understanding of the photophysical properties of squaraines at the example of N-isobutyl substituted anilino-squaraine (SQIB) by extending ground-state and excited-state absorption spectroscopy of the molecule into the ultraviolet up to 6.5~eV.
In addition, we distinguish the relative transition dipole moments of the excited state absorption peaks with the help of transient absorption anisotropy experiments.

To relate experimental features to specific states, we employ a set of \textit{ab initio} methods including time-dependent density functional theory (TDDFT), the Bethe-Salpeter equation (BSE) and n-electron valence perturbation theory on top of a self-consistent complete active space (CASSCF/NEVPT2). Our assignment is complemented by vibronic simulations and a discussion of two-photon absorption measurements.
Through this joint effort, we are able to provide a consistent picture of the optical behavior of SQIB across the visible and ultraviolet light regime, and assign a total of twelve electronically excited states to our experimental data.
\end{abstract}

\section{Introduction}

The molecular class of squaraine dyes is frequently discussed in the context of optoelectronic applications including organic photovoltaics\cite{Chen2015}, photosensors\cite{Csucker2025, Kim2021, Schulz2019, hu2024squaraine} or biomedical dyes\cite{He2020, Butnarasu2020, Hu2023sensor}.
The interest in these molecules stems from their appealing linear and non-linear optical properties.
Literature particularly highlights a sharp near-infrared optical absorption and fluorescence\cite{Zhao2024, Bondar2022} and a strong two-photon absorption cross
 section
 \cite{AvilaFerrer2019,Belfield2013,Basheer2007fluorescence,Ajayaghosh2005fluorescence}.
Advantageously, the molecules overall preserve these attributes also in the solid state with additional modifications arising from \textit{inter}molecular interactions \cite{Giavazzi2025a, Bernhardt2024, Balzer2022, Shen2021, Zheng2020, giavazzi2025modeling}.

\begin{figure}[h]
    \centering
    \includegraphics[width=0.5\linewidth]{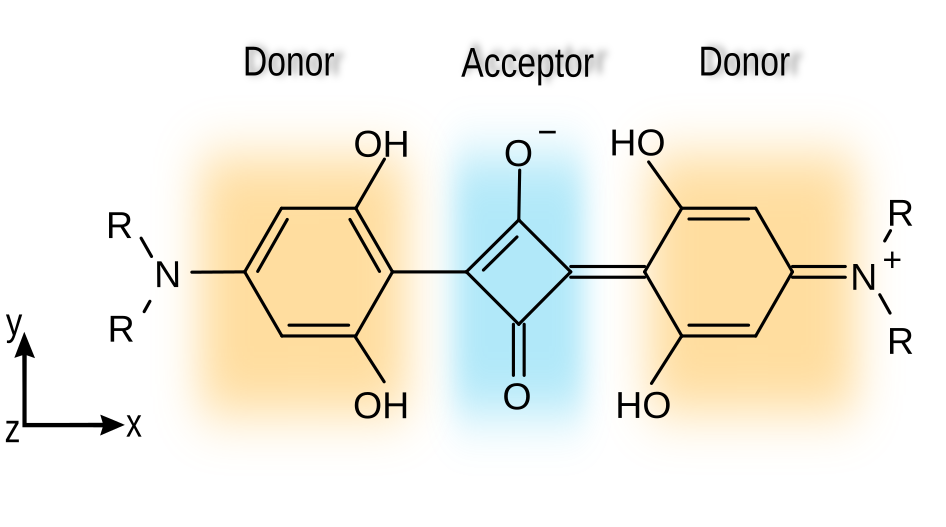}
    \caption{
    Schematic illustration of the molecular structure of anilino-squaraines. 
The electron-donating and electron-accepting regions are highlighted in orange and blue, respectively. 
In the experiments, SQIB (\(R =\mathrm{isobutyl}\)) was used, whereas for the calculations, the simplified model compound MeSQ (\(R =\mathrm{methyl}\)) was employed.
    }
    \label{fig:structure}
\end{figure}

To understand the origin of these characteristics, it is worth looking at the molecular level.
While nowadays there exists a wide variety of functional modifications to the group of squaraines\cite{Ilina2019, Xia2017, Beverina2014,shafeekh2013asymmetric} , here we focus on a classical centrosymmetric arrangement.
Specifically in our study, the name-giving squaric core is on both sides linked to a hydroxyanilino-ring (Fig.~\ref{fig:structure}).
The amino groups are saturated by isobutyl chains, resulting in isobutyl-anilino-squaraine (SQIB) in our case.
In such an arrangement, the electron-deficient center is usually attributed as an acceptor, compared to the more electron-rich aromatic side rings denoted as donors.
Together, they form the chromophoric system of the molecule. 

This donor-acceptor-donor type motif  (see Fig.~\ref{fig:structure}) is reflected by two degenerate zwitterionic resonance structures, where for instance the positive charge is located on one anilino nitrogen and the negative charge on one squaric oxygen. 
Fig.~\ref{fig:structure} also illustrates the quadrupolar nature of the molecule.
Alternatively, the molecule can also be described as a biradical with the two unpaired electrons on the squaric center.
How much these different charge-distributions actually contribute to the electronic states of the molecule, has not yet been fully resolved \cite{AvilaFerrer2019,prabhakar2005near,yesudas2007origin}. 
It is however without question that the unique structure is responsible for the electronic transitions.

In describing the optical properties, most investigations have concentrated on the visible regime ( 1.5 to 3.5 eV)\cite{webster2010near,scherer2002two, klishevich2025nonlinear}. 
From a practical point of view, this might simply be because many applications are primarily employing this energy region\cite{tao2020organic,li2013organic}. 
From a more fundamental perspective, it might be related to the challenges of interpreting experimental data.
Here, support from simulations of the electronic properties with \textit{ab initio} methods is notoriously difficult due to the intricate bonding situation of squaraines.
As Ferrer et al. \cite{AvilaFerrer2019} demonstrate in their seminal paper, up to date, a robust interpretation of the spectral features often requires a multitude of theoretical methods, which makes the analysis tedious and time-consuming.
In any case, the focus on the low-energy region has so far provided us with an in-depth understanding of mainly the first peak in one-photon and two-photon absorption of squaraine molecules.
The excited state landscape towards higher energies, however, is largely unexplored.

In our work, we aim to fill this gap by characterizing a broad energy range up to 6.5~eV.
To this end, we measured one-photon ground- and excited-state transmission spectra together with two-photon absorption spectra of isolated SQIB molecule. To interpret the spectral features, we combined our experimental results with calculations on the isolated MeSQ molecule, in which the alkyl substituent \( R \) (see Fig.~\ref{fig:structure}) was truncated to a methyl group. The calculations were performed at three different levels of theory: first, Casida's formulation of time-dependent density functional theory \cite{casida1995time} considering linear and quadratic response (TDDFT)\cite{casida2012progress,ullrich2012time}, second, the many-body perturbation framework of the Bethe-Salpeter equation on top of the GW approximation (GW/BSE) \cite{leng2016gw, onida2002electronic,blase2018bethe} and, third, the perturbative treatment of multi-configurational wave-functions within a complete active space (CASSCF/NEVPT2)\cite{king2022large,angeli2001introduction}.
Even though we employ three completely different theoretical approaches to compute optical spectra, we arrive at surprisingly similar interpretations.
All together, based on our data, we propose the assignment of twelve excited states in the studied energy window.
As an additional outcome, we provide a broad comparison between the three theoretical descriptions, which might guide future investigations into squaraines.

\section{Results and Discussion}

As the analysis in our study heavily depends on theoretical data, we start by clarifying some important settings in our calculations. 
First, we have chosen the squaraine molecule to be aligned with its long molecular axis along the $x$-direction of our coordinate system so that its aromatic backbone lies in the $xy$-plane as indicated in Fig.~\ref{fig:structure}. 
Furthermore, to make high-level theoretical calculations feasible, we replaced the isobutyl side chains of SQIB with methyl groups, yielding the smaller homologue MeSQ. 
This is a common approximation, as the alkyl substituents are not part of the chromophoric system and mainly influence molecular packing rather than the electronic transitions relevant in the studied energy range (see Fig.~\ref{fig:S2} in the Supporting Information). As an additional advantage, this increases the number of symmetry operations and allows us to classify the orbitals and electronic states of the molecule in the eight irreducible representations of the D$_{2h}$ point group.

\begin{figure}[h]
    \centering
    \includegraphics[width=0.5\linewidth]{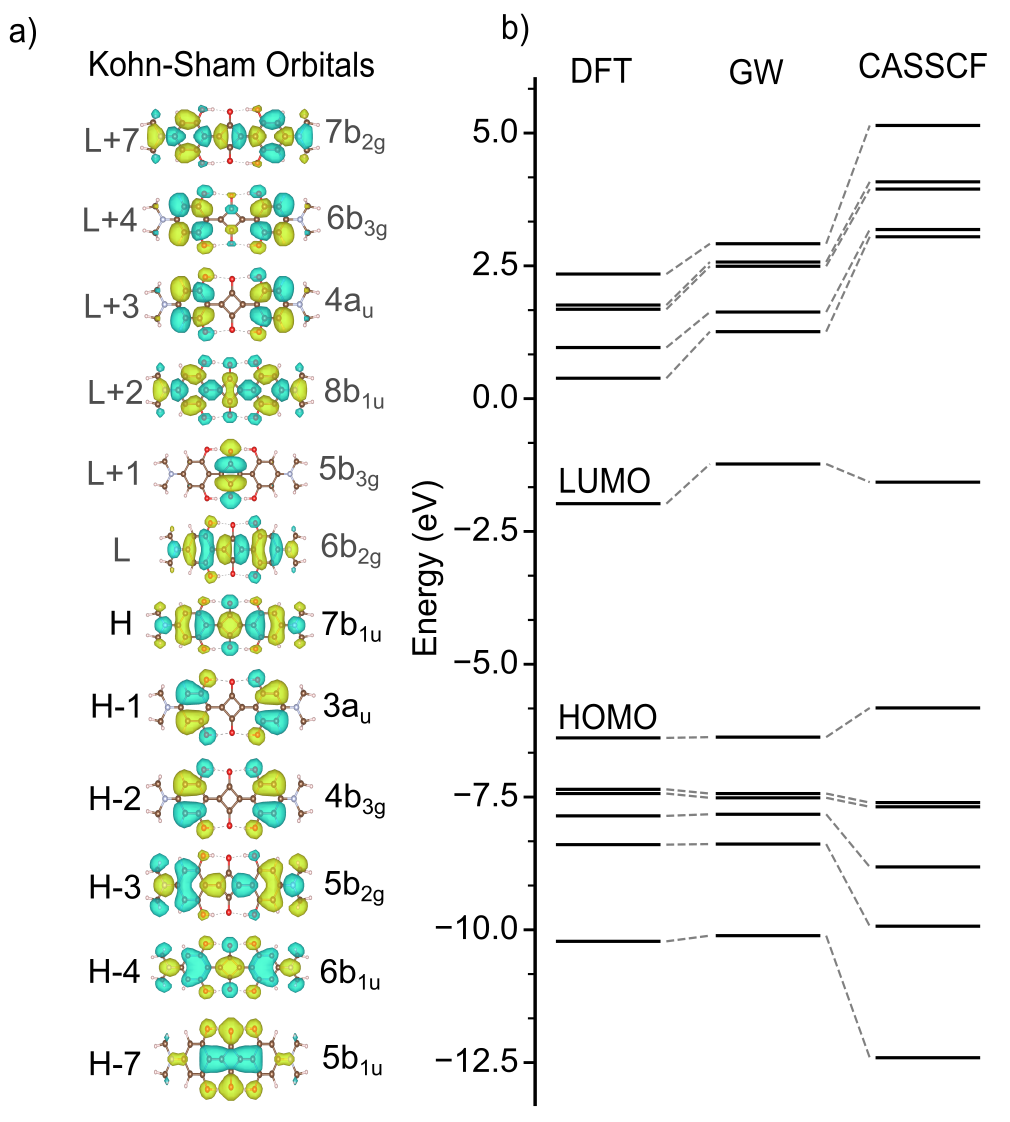}
    \caption{
   (a) Frontier $\pi$-orbitals of MeSQ with $D_{2h}$ symmetry (isovalue of 0.02) calculated at the DFT (CAM-B3LYP/cc-pVTZ) level. Each orbital is labeled according to its irreducible representation. Additionally, we label them by their energetic order in the DFT calculation as HOMO$-n$ and LUMO$+n$. Occupied orbitals are shown with black labels and unoccupied orbitals with grey labels. These molecular orbitals comprise the CAS(12,12) active space in our CASSCF calculations.  
(b) Comparison of orbital energy levels obtained from DFT, GW, and CASSCF calculations. Each horizontal line represents the energy of a corresponding orbital, and dashed lines connect orbitals of similar symmetry and character across the three computational methods.
    }
    \label{fig:orbitals}
\end{figure}

In our investigation, we have specifically concentrated on the frontier orbitals, which show $\pi$-character and are involved in the low-energy optical response.
Analyzing our TDDFT and GW/BSE calculations, we have found a set of 12 orbitals, namely the 6 highest occupied and 6 lowest unoccupied molecular $\pi$-orbitals, to be essential for capturing the excited states in the studied energy range.
Accordingly, we have also chosen these orbitals (Fig.~\ref{fig:orbitals}) as an active space in the multi-configurational calculations.
Besides the irreducible representation, we label the orbitals by their energetic ordering from DFT calculation counting the highest occupied (HOMO-n) and the lowest unoccupied molecular orbitals (LUMO+n). While the energetic order slightly varies with the level of theory, we observe no change in shape between the DFT and CASSCF-optimized orbitals. For more details, see Fig.~\ref{fig:S1}. 

The following discussion of the optical properties is divided into three parts. 
First, we focus on absorption from the ground state S$_0$ (GSA), then on absorption 
from the first excited state S$_1$ (ESA), and finally on the study of the 
two-photon absorption (2PA) process.

\subsection{Ground-State Absorption (GSA)}

\begin{figure}[!htbp]
    \centering
    \includegraphics[width=0.9\textwidth]{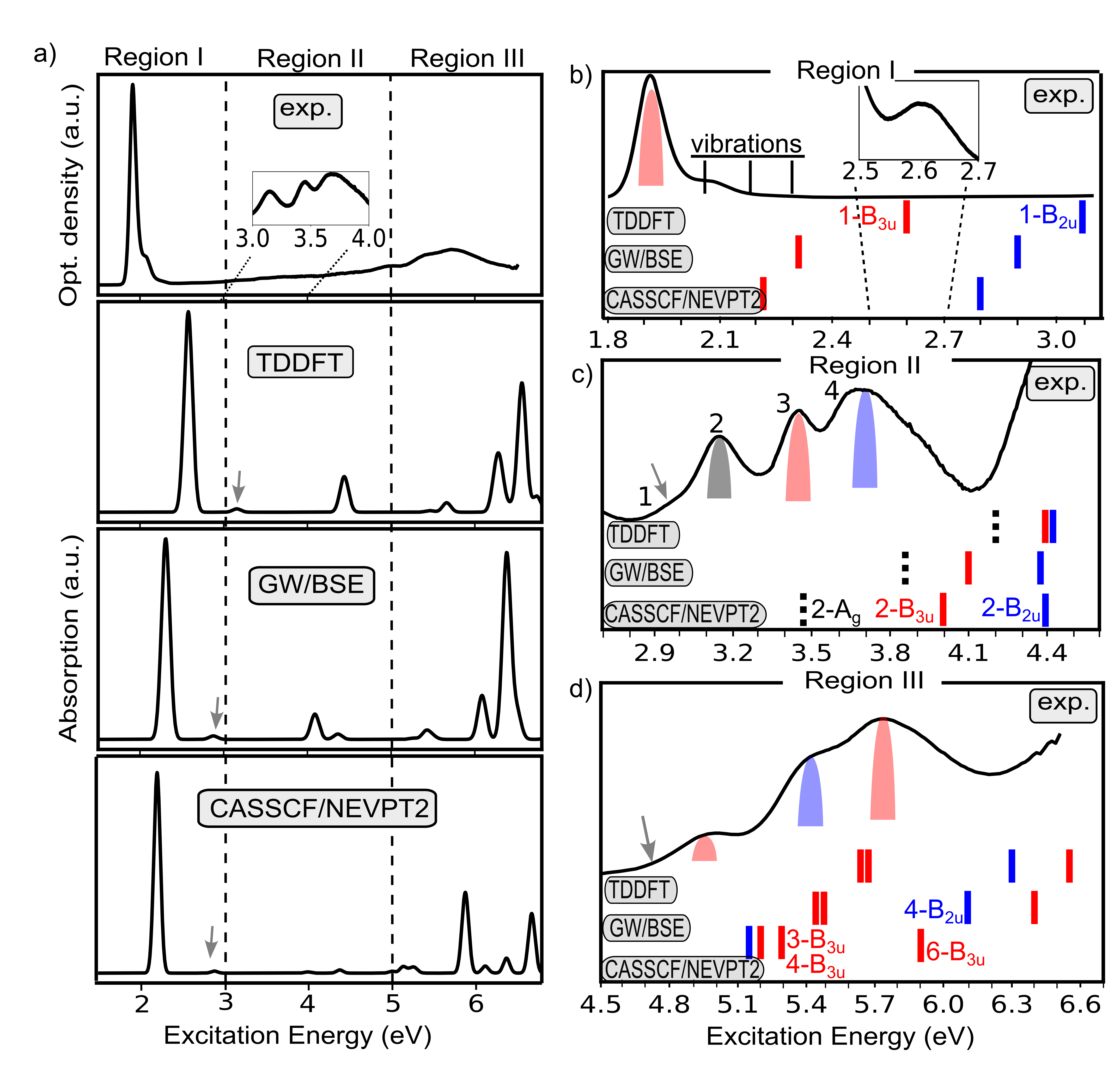}
    \caption{
 (a) Comparison of the experimental optical density of SQIB solvated in acetonitrile with the absorption spectrum of a single MeSQ molecule in the gas phase, calculated using TDDFT, GW/BSE, and CASSCF/NEVPT2. 
The top panel shows the experimental spectrum grouped into three spectral regions: I (1.5–3.0~eV), II (3.0–5.0~eV), and III (5.0–7.0~eV). 
The magnified inset highlights weak features in Region~II of the experimental spectrum. 
(b–d) Enlarged sections of the three energy regions showing the theoretical assignment of the experimental peak positions. 
Theoretical transition energies are indicated by colored bars, where red and blue correspond to transition dipoles polarized along the $x$- and $y$-directions, respectively. 
The grey arrows in panels~(a), (c), and~(d) mark weak peaks to make them more visible and to indicate the positions discussed in the text. 
    }
    \label{fig:gsa}
\end{figure}

Fig.~\ref{fig:gsa}(a) summarizes experimental and simulated GSA spectra for the isolated SQIB and MeSQ molecules, respectively. 
As an experimental reference, we used absorbance (= optical density) data measured for SQIB solvated in acetonitrile (top panel of Fig.~\ref{fig:gsa}(a)). This solvent was chosen due to its minimal spectral overlap with SQIB (cutoff~$<200$~nm).
To facilitate the discussion, the spectrum is grouped into three regions, denoted by the Latin numbers \RomanNumeralCaps{1}, \RomanNumeralCaps{2} and \RomanNumeralCaps{3}, respectively.

Region~I comprises the low-energy range (1.5-3.0~eV) and is dominated by an intense peak at 1.92~eV, which is characteristic for the first excited state in many anilino squaraines \cite{webster2010near,law1997absorption}.
In addition, our experiments resolve several vibronic progressions 
(for details, see Fig.\ref{fig:S4} and Fig.~\ref{fig:S5}), with the most pronounced feature 
at 2.07~eV, which corresponds to the symmetric stretching motion of the two anilino-groups relative to the squaric acid center~\cite{timmer2022charge}.

Region II ranging from 3.0-5.0~eV appears rather structureless at first glance, however, previous studies have located a weak electronic transition around 3.3~eV \cite{AvilaFerrer2019,webster2010near}. 
Indeed, the magnified inset in Fig.~\ref{fig:gsa} reveals three subtle peaks at 3.14, 3.45 and 3.68~eV in our experimental spectrum.

Finally, region~III spans the high-energy end of our experimentally accessible spectrum (5.0-7.0~eV).
This region features a broad absorption band without clearly resolved peaks, but visible shoulders at 5.03 and 5.48~eV as well as an absorption maximum at 5.75~eV. To the best of our knowledge, comparable experimental data have not been reported so far in literature.

The experimental spectrum is compared to the absorption spectra of a single MeSQ molecule in the gas phase computed at three different levels of theory, TDDFT, GW/BSE and CASSCF/NEVPT2, shown in the bottom panels of Fig.~\ref{fig:gsa}(a).
Before discussing the simulation results in more detail, it is worth recalling basic group theoretical considerations, which allow us to classify transitions as optically active.
According to previous literature, the S$_0$ ground state corresponds to the fully symmetric A$_g$ representation \cite{AvilaFerrer2019}. 
Moreover, in the $D_{2h}$ point group, the dipole operator for polarization along the $x$, $y$ and $z$-axis transforms according to $B_{3u}$, $B_{2u}$ and $B_{1u}$, respectively.
Consequently, for dipole-allowed transitions, an excited state also has to belong to either of the three symmetry representations with the polarization of the transition, then, being along the respective axis.

\begin{table*}[!ht]
    \centering
    \caption{
    Characterization of experimental and calculated peaks in the ground-state absorption spectrum of SQIB and MeSQ, respectively. 
    Experimental peak energies and relative peak heights (RPH) are compared with calculated excitation energies (in eV), oscillator strengths (\textit{f}), and dominant configuration interaction (CI) contributions for symmetry-allowed transitions in regions \RomanNumeralCaps{1}–\RomanNumeralCaps{3}, as obtained from TDDFT, GW/BSE, and CASSCF/NEVPT2 methods.
    Only dominant configurations with contributions above $5\%$ are included. 
    Molecular orbital labels correspond to those defined in Fig.~\ref{fig:orbitals}. 
    }
    \begin{adjustbox}{width=\textwidth}
    \begin{tabular}{@{} c c c c c l c l c l @{}}
    \toprule
        \multirow{3}[1]*{GSA}  & Exp.  & State.  & \multicolumn{2}{c}{TDDFT} & \multicolumn{2}{c}{GW/BSE} & \multicolumn{2}{c}{CASSCF/NEVPT2}\\
        & Energy & Sym & Energy & CI description & Energy & CI description & Energy & CI description\\
        & (RPH) & &(f) & & (f) & & (f) & \\
        \cmidrule(lr){4-5}\cmidrule(lr){6-7}\cmidrule(lr){8-9}

    \midrule
    \vspace{0.5em}
        \multirow{2}*{I} & \begin{tabular}[t]{c}1.92 \\(1.00) \end{tabular} & 1-B$_{3u}$ &
        \begin{tabular}[t]{c} 2.58 \\ (1.60)\end{tabular} & 
        7$b_{1u}\to$6$b_{2g}$ (96\%) & 
        \begin{tabular}[t]{c} 2.31 \\ (1.33)\end{tabular}& 
        7$b_{1u}\to$6$b_{2g}$ (99\%) & 
        \begin{tabular}[t]{c} 2.20 \\ (1.63)\end{tabular} & 
        7$b_{1u}\to$6$b_{2g}$ (72\%) \\
    \midrule
        \multirow{10}*{II}   &  \begin{tabular}[t]{c}3.14 \\(0.02) \end{tabular}   &  2-A$_{g}$ & 
        \begin{tabular}[t]{c} 4.21 \\ (--)\end{tabular} &
        5$b_{2g}\to$6$b_{2g}$ (94\%) & 
        \begin{tabular}[t]{c} 3.84 \\ (--)\end{tabular} &
        5$b_{2g}\to$6$b_{2g}$ (98\%) & 
        \begin{tabular}[t]{c} 3.48 \\ (--)\end{tabular} & 
         \begin{tabular}[t]{l}
      7$b_{1u}\to$6$b_{2g}$ (2e) (34\%) \\
      5$b_{2g}\to$6$b_{2g}$ (34\%) \\
      7$b_{1u}\to$8$b_{1u}$ (11\%)
         \end{tabular}\\
    \cmidrule{2-9}
        &  \begin{tabular}[t]{c}3.45 \\(0.03) \end{tabular}   &  2-B$_{3u}$ &
        \begin{tabular}[t]{c} 4.44 \\ (0.25)\end{tabular} &
        6$b_{1u}\to$6$b_{2g}$ (94\%) & 
        \begin{tabular}[t]{c} 4.10 \\ (0.17)\end{tabular} & 
        6$b_{1u}\to$6$b_{2g}$ (94\%) & 
        \begin{tabular}[t]{c} 4.00 \\ (0.01)\end{tabular} & 
        \begin{tabular}[t]{l}
      6$b_{1u}\to$6$b_{2g}$ (45\%) \\
      7$b_{1u}$,5$b_{2g}\to$6-$b_{2g}$ (14\%) \\
      7$b_{1u}\to$8$b_{1u}$,6$b_{2g}$ (7\%)
        \end{tabular}\\
    \cmidrule{2-9}
    \vspace{0.5em}
        &  \begin{tabular}[t]{c}3.68 \\(0.04) \end{tabular}   &  2-B$_{2u}$ &
        \begin{tabular}[t]{c} 4.47 \\(0.04)\end{tabular} & 
        7$b_{1u}\to$5$b_{3g}$ (95\%) & 
        \begin{tabular}[t]{c} 4.38 \\ (0.04)\end{tabular} &
        7$b_{1u}\to$5$b_{3g}$ (97\%) & 
        \begin{tabular}[t]{c} 4.39 \\ (0.02)\end{tabular} &
        7$b_{1u}\to$5$b_{3g}$ (74\%) \\
    \midrule
\multirow{12}*{III} & \begin{tabular}[t]{c}5.03 \\(0.09) \end{tabular} & 3-B$_{3u}$ &
  \begin{tabular}[t]{@{}c@{}} 5.66 \\ (0.06) \end{tabular} &
  \begin{tabular}[t]{@{}l@{}} 3$a_u\to$5$b_{3g}$ (93\%) \end{tabular} &
  \begin{tabular}[t]{@{}c@{}} 5.44 \\ (0.06) \end{tabular} &
  \begin{tabular}[t]{@{}l@{}} 5$b_{1u}\to$6$b_{2g}$ (96\%) \end{tabular} &
  \begin{tabular}[t]{@{}c@{}} 5.22 \\ (0.02) \end{tabular} &
\begin{tabular}[t]{@{}l@{}}
3$a_u \to 5b_{3g}$ (60\%)\\
7$b_{1u}$, 4$b_{3g}$ $\to$ 6$b_{1u}$, 5$b_{3g}$ (9\%)\\
\end{tabular} \\

\cmidrule(lr){3-9} 
 &       &  4-B$_{3u}$ &
  \begin{tabular}[t]{@{}c@{}} 5.67 \\ (0.02) \end{tabular} & 
  \begin{tabular}[t]{@{}l@{}} 5$b_{1u}\to$6$b_{2g}$ (90\%) \end{tabular} &
  \begin{tabular}[t]{@{}c@{}} 5.49 \\ (0.02) \end{tabular} &
  \begin{tabular}[t]{@{}l@{}} 3$a_u\to$5$b_{3g}$ (98\%) \end{tabular} &
  \begin{tabular}[t]{@{}c@{}} 5.28 \\ (0.04) \end{tabular} &
\begin{tabular}[t]{@{}l@{}}
5$b_{1u} \to $6$b_{2g}$ (26\%)\\
7$b_{1u} \to $7$b_{2g}$ (15\%)
\end{tabular}
\\
\cmidrule(lr){2-9}
& \begin{tabular}[t]{c} 5.48 \\ (0.16) \end{tabular} & 4-B$_{2u}$ &
\begin{tabular}[t]{c} 6.30 \\ (0.30) \end{tabular} &
\begin{tabular}[t]{l}
4$b_{3g}\to$8$b_{1u}$ (53\%)\\
7$b_{1u}\to$6$b_{3g}$ (28\%)
\end{tabular} &
\begin{tabular}[t]{c} 6.10 \\ (0.29) \end{tabular} &
\begin{tabular}[t]{l}
4$b_{3g}\to$8$b_{1u}$ (56\%)\\
7$b_{1u}\to$6$b_{3g}$ (30\%)
\end{tabular} &
\begin{tabular}[t]{c} 5.14 \\ (0.05) \end{tabular} &
\begin{tabular}[t]{l}
7$b_{1u}\to$6$b_{3g}$ (35\%)\\
7$b_{1u}$, 4$b_{3g}\to$6$b_{2g}$ (12\%)\\
5$b_{2g}$,3$a_{u}\to$6$b_{2g}$ (10\%)
\end{tabular} \\
    \cmidrule{2-9}
        &  \begin{tabular}[t]{c}5.75 \\(0.18) \end{tabular}   &   6-B$_{3u}$ &
        \begin{tabular}[t]{c} 6.56 \\ (1.03)\end{tabular} &
        \begin{tabular}[t]{c}
        7b$_{1u}\to$7b$_{2g}$ (28\%) \\
        3a$_u\to$6b$_{3g}$ (18\%) \\
        4b$_{3g}\to$4a$_u$ (18\%) \\
        5b$_{2g}\to$8b$_{1u}$ (9\%)
        \end{tabular} &
        \begin{tabular}[t]{c}6.38 \\ (1.23) \end{tabular} &
        \begin{tabular}[t]{c}
        4b$_{3g}\to$4a$_u$ (38\%) \\
        3a$_u\to$6b$_{3g}$ (37\%) \\
        5b$_{2g}\to$8b$_{1u}$ (11\%) \\
        7b$_{1u}\to$7b$_{2g}$ (7\%)
        \end{tabular} & 
        \begin{tabular}[t]{c}5.87 \\ (0.21) \end{tabular} &
        \begin{tabular}[t]{c}
        4b$_{3g}\to$4a$_u$ (18\%) \\
        5b$_{2g}\to$8b$_{1u}$ (16\%) \\
        3a$_u\to$6b$_{3g}$ (14\%) \\
        5b$_{1u}\to$6b$_{2g}$ (14\%)
        \end{tabular} \\        
    \bottomrule
    \end{tabular}
    \label{tab:gsa}
    \end{adjustbox}
\end{table*}

Table~\ref{tab:gsa} summarizes the energies, oscillator strengths, and main electronic configurations of the excited states responsible for the most prominent features in each region.
A more detailed summary listing all symmetry-allowed transitions, which we have calculated for MeSQ, is provided in the SI,  Table~\ref{table:S1}.
Note that in our multi-configurational approach, we have not considered excitations along the $z$-direction (B$_{1u}$).
In the considered energy range, those transitions are expected to have negligible contributions, since the chromophoric system is flat and lies in the $x$–$y$ plane.
 
In general, our TDDFT and GW/BSE calculations support this assumption.

As a reference to previous investigations, we start by discussing data for the most prominent excitation of the spectrum at 1.92~eV in region \RomanNumeralCaps{1} (Fig.~\ref{fig:gsa}(b)).
Consistent with the references\cite{AvilaFerrer2019,timmer2022charge}, we assign it to the first excited state S$_1$, which has B$_{3u}$ symmetry (1-B$_{3u}$), and predominantly arises from a HOMO (7b$_{1u}$) $\rightarrow$ LUMO (6b$_{2g}$) transition with weights of 96\% (TDDFT), 99\% (GW/BSE), and 72\% (CASSCF/NEVPT2). Contributions from all other transitions are below 5\%.

When comparing the calculated with the experimental peak position at 1.92~eV, we observe an increasing improvement when climbing the theoretical ladder. While TDDFT (2.58~eV) clearly overestimates the experimental excitation energy (1.92~eV) by 0.61~eV, GW/BSE approaches the experimental value already closer (2.31~eV). Finally, CASSCF/NEVPT2 locates the peak at 2.20~eV.

In this regard, it should be noted that the presented calculations do not account for spectral shifts arising from solvent effects. 
While a rigorous treatment of the solvent environment was not feasible at all levels of theory, 
an initial estimate was obtained using conductor-like polarizable continuum modeling (CPCM) within DFT (see Fig.~\ref{fig:S3}). 
At this level, the first excited state exhibits a red-shift of approximately 0.2\,eV when comparing gas-phase to solvent-phase calculations.
Assuming a shift of a similar magnitude for the CASSCF/NEVPT2 calculation, brings our result in perfect agreement with experiment.

Completing our investigation of the first peak, we have also computed the vibrational properties of MeSQ to account for the vibronic progression in the experimental data. 
Both, Franck--Condon and Herzberg--Teller contributions were evaluated within the ``Adiabatic Hessian After a Step'' approximation at the DFT level, yielding results consistent with previous theoretical reports~{\cite{timmer2022charge,burigana2025signatures} {for more details, see Fig.~\ref{fig:S4} and Fig.~\ref{fig:S5}).
In Region~I, vibronic analysis of the first excited state (S$_1$) shows that the spectral features are mainly governed by Franck--Condon transitions, with Huang--Rhys factors and dimensionless displacements identifying the dominant modes, as shown in Fig.~\ref{fig:S4}. The vibrational analysis is consistent with previous observations \cite{timmer2022charge, burigana2025signatures}. 

Another peak, featuring a considerably smaller oscillator strength compared to the bright $B_{3u}$ state, can be found at the upper edge of region~I at 2.61~eV (see inset in Fig.~\ref{fig:gsa}(b)).
From our vibrational analysis of Fig.~\ref{fig:S4}, we are not able to explain this feature with a vibronic progression of the $S_1$.
Instead, we relate it to the first excitation polarized along the shorter molecular $y$-axis, labeled 1-$B_{2u}$.
In our calculations, the state is described by an excitation from the HOMO-1 (3$a_u$) to the LUMO (6$b_{2g}$). 
Depending on the method, its energy position shifts from 3.16~eV (TDDFT) to 2.88~eV (GW/BSE) and 2.79~eV (CASSCF/NEVPT2), as indicated by the grey arrow in Fig.~\ref{fig:gsa}(a) showing the position of this minor peak in the calculations.
 
Moving on to region~\RomanNumeralCaps{2}, we address the four subtle signals around 3.5~eV (Fig.~\ref{fig:gsa}(c)). 
The interpretation of those proves to be intricate. 
Although our simulations locate in total four possible states in this energy range, only two of those are inherently symmetry-allowed. 
Firstly, there is the second state of $B_{3u}$ symmetry, 2-$B_{3u}$, which mainly arises from a HOMO-4 (6$b_{1u}$) to LUMO (6$b_{2g}$) transition, admixed with minor double-excitations in the multi-configurational approach. 
Secondly, the neighboring state of $B_{2u}$ symmetry, 2-$B_{2u}$, involves a transition between HOMO (7$b_{1u}$) and LUMO+1 (5$b_{3g}$). 
Their assignment to experimental features is complicated by an larger offset of the calculated excitation energies than encountered in region~\RomanNumeralCaps{1}.
TDDFT and GW/BSE places the states at 4.44 and 4.47~eV (error of 0.8-1.2~eV) and 4.10 and 4.38~eV (error of 0.5-0.9~eV), respectively.
With CASSCF/NEVPT2, the states shift down to 4.00 and 4.39~eV. 
We estimate additional 0.09~eV from solvent effects, but still, the excitations are predicted notably too high in energy to unambiguously identify them as two of the experimental maxima (3.14, 3.45, 3.68~eV).

Thus, we are left with a tentative assignment based on oscillator strength, energy and especially the ordering of states in our simulations.
Across all three methods, the  2-$B_{3u}$ and 2-$B_{2u}$ are placed energetically above the other two states in region ~\RomanNumeralCaps{2}, to which the transitions are dipole-forbidden.
Having this in mind, we assign the states to the experimental maxima at higher energy, namely the peak at 3.45\,eV to the 2-B$_{3u}$ state and the peak at 3.68\,eV to the 2-B$_{2u}$ state.
Interestingly, the relative intensity between the 2-B$_{3u}$ and 2-B$_{2u}$ states changes across the theoretical methods. 
While TDDFT predicts a considerably stronger 2-B$_{3u}$ transition, CASSCF/NEVPT2 yields comparable oscillator strengths for both states, indicating a relative enhancement of the higher-lying 2-B$_{2u}$ state. 
This trend aligns well with the experimental observation, where the higher-energy feature exhibits greater intensity.
Our assignment is further strengthened by our vibronic calculations shown in Fig.~\ref{fig:S5}. 
While the 2-B$_{3u}$ state is predicted as a sharp absorption feature, the vibrational contributions to the 2-B$_{2u}$ broaden the transition in energy explaining the high-energy tail in the experimental data.

Regarding the peak at 3.14~eV, we suspect its origin in one of the two dark states lying below the 2-$B_{3u}$ state.
Note that even though excitations are symmetry-forbidden, they could borrow intensity through the Herzberg-Teller effect \cite{lin1974study,nooijen2006investigation}.
This assumption is at least partially supported by vibrational calculations, where we indeed find that the 2-$A_g$ transition gains intensity from vibrational contributions (see Fig.\ref{fig:S5}). In addition, the minor experimental feature at 2.97~eV (marked by a grey arrow in Fig.~\ref{fig:gsa}(c) can be assigned to the 1-B$_{3g}$ state, which likewise gains weak intensity through Herzberg--Teller coupling in our vibrational analysis.
Note that 1-$A_g$ denotes the ground state.
With an energy difference of 0.23~eV (TDDFT), 0.26~eV (GW/BSE) and 0.52~eV (CASSCF/NEVPT2) relative to the symmetry-allowed 2-B$_{3u}$ transition, the 2-$A_g$ position would be in the range of the experimental difference ($\Delta E_{\text{peak-2}\rightarrow\text{peak-3}} = 0.31~\text{eV}$).
However, we also admit that the calculated vibronic intensity gain amounts to only a fraction of what is observed experimentally. 

Finally, we discuss region~\RomanNumeralCaps{3} (Fig.~\ref{fig:gsa}d). 
In this range, our theoretical results suggest a dense manifold of electronic transitions with mixed orbital characters. 
In the following, we highlight a few of those states, which we regard as important contributions to the experimental spectrum.

Our three methods coincide in predicting an intense peak, 6-$B_{3u}$, located at 6.56~eV by TDDFT, at 6.38~eV by GW/BSE and at 5.87~eV by CASSCF/NEVPT2.
Accounting for an approximate shift of 0.15~eV due to the solvent environment, we attribute this peak to the maximum of the experimental signal at 5.75~eV.
While all three methods agree on this feature in general, its nature is less clear since various single-particle transitions contribute.
At least, the description at all three levels shares a mixing of two transitions, namely between the HOMO-2 (4$b_{3g}$) to LUMO+3 (4$a_u$) and the HOMO-1 (3$a_u$) to LUMO+4 (6$b_{3g}$).

As a second common prediction, all calculations show a significantly weaker signal 0.89 (TDDFT), 0.93 (GW/BSE) and 0.62~eV (CASSCF/NEVPT2) below the strong 6-$B_{3u}$ absorption peak.
It consists of two $B_{3u}$ excitations, 3-$B_{3u}$ and 4-$B_{3u}$ placed closed to each other at about 5.67 (TDDFT), 5.49 (GW/BSE) and 5.28~eV (CASSCF/NEVPT2). 
Therefore, they are most likely responsible for the shoulder at 5.03~eV ($\Delta$E = 0.72~eV to experimental maximum) at the onset of region~\RomanNumeralCaps{3}.
For all methods, one of the two states shows major contribution from a HOMO-7 (5$b_{1u}$) to LUMO (6$b_{2g}$) transition.
Its counterpart is mainly described by an excitation from the HOMO-1 (3$a_u$) to the LUMO+1 (5$b_{3g}$).
Interestingly, their relative intensities depend on the level of theory, with TDDFT and GW/BSE favoring the first peak rather than the second as in CASSCF/NEVPT2.

Having assigned the peaks at both the lower and upper ends of the absorption band, we turn our attention to the shoulder observed at 5.48~eV. 
Based on our analysis, this feature should be attributed to a state lying between the previously discussed B$_{3u}$ excitations. All three theoretical methods -- TDDFT, GW/BSE, and CASSCF/NEVPT2 -- consistently predict the 4-B$_{2u}$ transition in this energy range, albeit only TDDFT and GW/BSE give it a comparable intensity.
This discrepancy is also reflected in the excitation-contributions, where the main HOMO-2 (4b$_{3g}$)~ to ~LUMO+2 (8b$_{1u}$) transition is only marginal in CASSCF/NEVPT2.
The methods agree, however, on a 30\% excitation from the HOMO (7b$_{1u}$)~ to the LUMO+4 (6b$_{3g}$)

To conclude this first section, we find a qualitatively similar description of the ground state absorption by all three tested methods. 
The most prominent features are related to the same single-particle transitions, showing agreement not only in the electronic configurations but also in the relative oscillator strengths.
Thus, important for the understanding of the absorption process, we would arrive at the same interpretation of the GSA spectrum irrespective of the chosen method. 
Regarding the energetics, a systematic shift toward the experimental values with higher-level theory was to be expected and simply highlights the influence of electron correlation for excited states.
\subsection{Excited-State Absorption (ESA)}

Owing to dipole selection rules, the ground state absorption spectrum has offered insights into the $B_{3u}$ and $B_{2u}$ excited states of isolated N-alkylated anilino Squaraines (SQIB) and MeSQ.
Further insight into the excited state landscape of the molecules can be gained by turning to two-photon absorption processes.
We start by performing pump-probe experiments 
by, first, optically pumping the intense transition from the ground state to the first excited state 1-$B_{3u}$, and then probe the optical absorption of the transiently excited molecular species. 
In this way, we have access to an entirely different set of excited states.
Considering the symmetry of the transition dipole operator ($B_{3u}$, $B_{2u}$, $B_{1u}$), for squaraines with $S_1$ showing B$_{3u}$ character, excitations into states of $A_g$, $B_{1g}$ and $B_{2g}$ symmetry, respectively, are possible.
Here, $A_g$ states are of special interest, since their transition dipoles are polarized along the $x$-direction and expected to yield the strongest excited state absorption peaks.

\begin{figure*}[t]
    \centering \includegraphics[width=0.6\textwidth]{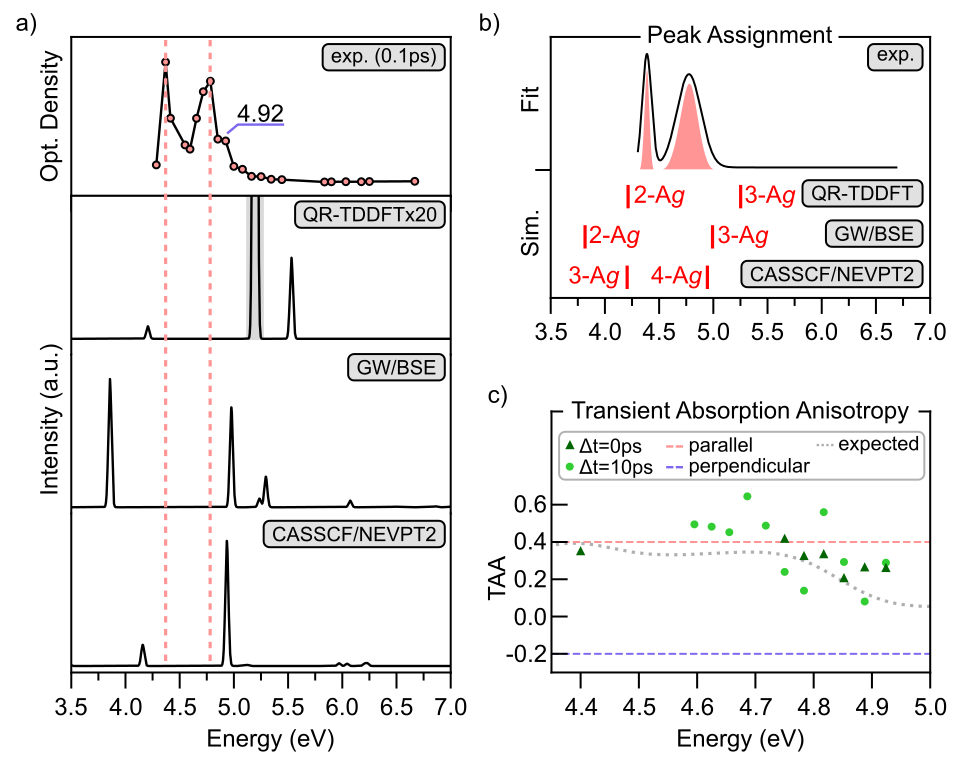}
\caption{
(a) Comparison between the experimental transient absorbance spectra of SQIB molecules dispersed in a PMMA matrix. Experimental spectra recorded at a 0.1 ps delay time between the pump and probe pulses and the theoretical spectra of an isolated MeSQ molecule in the gas phase, obtained from QR-TDDFT, GW/BSE, and CASSCF/NEVPT2 calculations, respectively. 
The zero of the energy scale corresponds to the ground state, $S_0$. 
The QR-TDDFT spectrum is scaled by a factor of 20 for clarity. 
(b) Gaussian fit of the experimental ESA signal, overlaid with the calculated excitation energies from QR-TDDFT, GW/BSE, and CASSCF/NEVPT2. 
Theoretical assignments of the main transitions are indicated by red bars, where $x$- and $y$-polarized transition dipoles are represented by red and blue labels, respectively. 
(c) Transient absorption anisotropy at $\Delta t = 0$~ps (dark-green triangles) and $\Delta t = 10$~ps (light-green circles) as a function of absolute energy. Both delay times correspond to independent experimental data sets, illustrating the overall trend rather than exact quantitative values. The dashed red and purple lines indicate the limiting cases of purely parallel and purely perpendicular ESA transition dipole moments with respect to the pump--probe polarization direction. 
}
    \label{fig:esa}
\end{figure*}

Fig.~\ref{fig:esa}(a) presents a comparison between the experimental excited-state absorption spectrum of SQIB and theoretical results obtained from QR-TDDFT, GW/BSE, and CASSCF/NEVPT2 calculations of MeSQ (for further details, see Method Section). 
The spectra are plotted as a function of the absolute transition energy, corresponding to the sum of the pump and probe photon energies. This is justified for SQIB due to the small Stokes shift of approximately 30~meV in solution \cite{Wang2011emission}, which would only result in a slightly blue-shifted ESA spectrum, and due to the absence of unoccupied electronic states below the pump energy, preventing population transfer processes.
The absolute transition energy facilitates the comparison to the excited state energies discussed in the previous section.
The experimental spectrum in the top panel of Fig.~\ref{fig:esa}(a) was obtained from transient absorption measurements on 1\% of SQIB molecules in a PMMA matrix using parallel pump and probe polarizations, where the pump beam is fixed to a central wavelength of 653 nm, i.e.: to the 1.9 eV $S_1$ state. 
As shown in Fig.~\ref{fig:S6}, the normalized GSA spectra of SQIB in chloroform and embedded in a PMMA matrix exhibit almost identical spectral shapes and peak positions, indicating that the two environments affect the ground-state electronic transitions in similar (for consistency between the GSA and ESA analyses, Fig.~\ref{fig:S6}(a) also shows GSA spectra of SQIB in chloroform and acetonitrile. 
These data demonstrate that SQIB exhibits negligible solvatochromism). 
We expect analogue behavior for the excited-state absorption spectra. 

Concerning the measured spectrum, we want to emphasize that there are only two prominent features in the entire experimentally accessible window between 3 and 7~eV, namely at 4.38~eV and 4.78~eV.
Their energy positions and relative intensities are in good agreement with previous reports\cite{zheng2020measurement}.
In addition, our experimental data suggests a weak third contribution at 4.93~eV as a high-energy shoulder of the second peak.
Note that the absence of absorption between 3.5 and 4~eV may be due to ground state depletion effects (1.9~eV "pump" excitation to $S_1$" and 1.9~eV "probe"), which precludes more definite statements about this energy range.

Turning to the comparison with our theoretical results, we find a clear contrast to the case of ground state absorption, as the three theoretical spectra differ by more than a mere energy shift.
This does not come unexpected since predicting especially $A_g$ states is known as a tough theoretical challenge due to the prominent role of double-excitations.
Frameworks like TDDFT or GW/BSE, which base their description of excited states on single-particle transitions, inherently lack such effects making $A_g$ states ill-described.
This complicates not only the comparison to experiment but also among the three theoretical approaches.
Despite the obvious differences, there are also a number of similarities that we highlight in the following.

\begin{table*}[ht]
\centering
\caption{
Comparison of experimental and theoretical ESA transitions of SQIB and MeSQ, respectively. 
Experimental peak positions are compared with calculated transition energies and oscillator strengths (in parentheses) obtained using QR-TDDFT, GW/BSE, and CASSCF/NEVPT2 methods. 
The dominant configuration interaction (CI) contributions (above $5\%$) and corresponding excited-state symmetries are listed for each transition.
}
\begin{adjustbox}{width=\textwidth}
\begin{tabular}{ccclclcl}
\toprule
    \multirow{2}{*}{ESA}  & Exp. & \multicolumn{2}{c}{QR-TDDFT} & \multicolumn{2}{c}{GW/BSE} & \multicolumn{2}{c}{CASSCF/NEVPT2}\\
    & Energy & Energy & \multirow{2}{*}{CI description / sym} & Energy & \multirow{2}{*}{CI description / sym} & Energy & \multirow{2}{*}{CI description / sym}\\
    & (f) & (f) & & (f) & & (f) & \\
\midrule
\vspace{0.5em}
    I & 4.38 & 
    \begin{tabular}[t]{c} 4.20 \\ (0.08) \end{tabular} & 
    \begin{tabular}[t]{c}
        5b$_{2g}\to$6b$_{2g}$ (96\%) \\
        2-A$_{g}$
    \end{tabular} & 
    \begin{tabular}[t]{c} 3.85 \\ (1.01) \end{tabular} & 
    \begin{tabular}[t]{c}
        5b$_{2g}\to$6b$_{2g}$ (98\%) \\
        2-A$_{g}$
    \end{tabular} & 
    \begin{tabular}[t]{c} 4.15 \\ (0.36) \end{tabular} & 
    \begin{tabular}[t]{c}
        5b$_{2g}\to$6b$_{2g}$ (35\%) \\
        7b$_{1u}\to$8b$_{1u}$ (11\%) \\
        3-A$_{g}$
    \end{tabular} \\
\midrule
    \multirow{1}{*}{II} & 4.78 & 
    \begin{tabular}[t]{c} 5.19 \\ (12.49) \end{tabular} & 
    \begin{tabular}[t]{c}
        7b$_{1u}\to$8b$_{1u}$ (82\%) \\
        3-A$_{g}$
    \end{tabular} & 
    \begin{tabular}[t]{c} 4.97 \\ (0.79) \end{tabular} & 
    \begin{tabular}[t]{c}
        7b$_{1u}\to$8b$_{1u}$ (93\%) \\
        3-A$_{g}$
    \end{tabular} & 
    \begin{tabular}[t]{c} 4.93 \\ (2.46) \end{tabular} & 
    \begin{tabular}[t]{c}
        7b$_{1u}\to$6b$_{2g}$ (2e) (29\%) \\ 7b$_{1u}\to$8b$_{1u}$ (27\%) \\
        4-A$_{g}$
    \end{tabular} \\
\bottomrule
\end{tabular}
\end{adjustbox}
\label{tab:esa}
\end{table*}

In agreement with experiment, all three levels of theory suggest strong excited state absorption peaks in the range of 4 to 5~eV, which are listed in Table~\ref{tab:esa}.
While the GW/BSE and CASSCF/NEVPT2 spectra display nuanced structures, the QR-TDDFT spectrum is dominated by a very intense single peak (shaded area in Fig.~\ref{fig:esa}(a)).
However, it is important to note that the oscillator strength of this dominant transition (transition dipole moment exceeding 25~\AA) is unrealistically large for a single molecule on the length scale of our squaraine (13~\AA). 

Such an artificial enhancement is a known limitation of quadratic-response TDDFT, occurring whenever the transition energy between two excited states becomes nearly resonant with another excitation.
In our case, QR-TDDFT predicts excited state absorption into a state 2.61~eV above the $S_1$, which coincides with excitation of the $S_1$ itself (2.58~eV in TDDFT). 
Under these conditions, spurious poles in the QR equations lead to an overestimated orbital relaxation term and unphysical peak intensity.
\cite{dalgaard1982quadratic, parker2016unphysical, parker2018quadratic}. 
Counteracting this behavior, we have magnified the QR-TDDFT results in Fig.~\ref{fig:esa}(a) and, thus, include seemingly smaller peaks in our discussion.

Regarding the first experimental ESA signal at 4.38~eV, we may compare it to the first appreciable peak in the simulations. 
For QR-TDDFT and GW/BSE, this would correspond to the 2-$A_g$ state, which is predicted to be at 4.20 and 3.85~eV, respectively.
Note that in our GW/BSE approach, this feature even shows the strongest ESA intensity.
However, we have already discussed this state together with the GSA results of region \RomanNumeralCaps{2}, where we suggested it to be vibrationally enhanced and the origin of a shoulder around 3~eV.
As we will discuss below, two-photon absorption measurements further strengthen our original assignment for the  2-$A_g$ to be at 3~eV.

Turning to CASSCF/NEVPT2 instead reveals the first ESA active state to actually be the 3-$A_g$, which is calculated to be at 4.15~eV. 
Interestingly, its major contributions are the HOMO-3 (5$b_{2g}$) and LUMO (6$b_{2g}$), similar to the 2-$A_g$ state of QR-TDDFT and GW/BSE.

In contrast, the 2-$A_g$ in the multi-configurational approach shows strong double-excitation character and therefore has no counterpart in the QR-TDDFT or GW/BSE description. Furthermore, CASSCF/NEVPT2 contributes only minor ESA intensity to the 2-$A_g$, so that in the context of our experimental data, a corresponding signal is either too low or covered by the ground-state depletion feature (see SI, Fig.~\ref{fig:S7}).

In order to rationalize the population of this highly excited state in terms of a pump-probe excitation process, we recall the $S_0\rightarrow S_1$ excitation, which is initiated here by the first laser pulse, populating the 1-$B_{3u}$ state through HOMO (7$b_{1u}$) to LUMO (6$b_{2g}$) transition (see Tab.~\ref{tab:gsa}). 
This shows that the second laser pulse excites the 3-$A_g$ state through transition from HOMO-3 (5$b_{2g}$) to the depopulated HOMO (7$b_{1u}$) orbital. 
Note that the $B_{3u}$ symmetry of this second transition predicts that the corresponding dipole moment is oriented along the long molecular axis, as for the first transition.  
The polarizations of the first and second laser pulse have thus to be parallel for maximum ESA signal.  

For the interpretation of the second peak at 4.78~eV, we encounter a similar situation.
In QR-TDDFT and GW/BSE, the second absorption with appreciable oscillator strength belongs to the 3-$A_g$ state at 5.19~eV and 4.97~eV, respectively. 
 
This state is mainly comprised of a single-particle transition from the HOMO (7$b_{1u}$) to the LUMO+2 (8$b_{1u}$).
An equivalent state in CASSCF/NEVPT2, albeit mixed with major double-excitation contributions, is located at 4.93~eV and is labeled the 4-$A_g$.
In a pump-probe excitation process with $S_1$ population as the first step (see Tab.~\ref{tab:gsa}), the second laser pulse excites the 4-$A_g$ state through transition from LUMO (6$b_{2g}$) to the LUMO+2 (8$b_{1u}$) orbital. 
This second transition has again $B_{3u}$ symmetry, requiring parallel polarization of the first and second laser pulse along the long molecular axis.

In summary, QR-TDDFT and GW/BSE would relate the two ESA features to first two $A_g$ excited states (2-$A_g$ and 3-$A_g$).
Our CASSCF/NEVPT2 calculation likewise finds those consecutive states, however denotes them as 3-$A_g$ and 4-$A_g$ due to the presence of a further state at lower energies described by double-excitations.
We conclude that even though QR-TDDFT and GW/BSE bear some apparent disadvantages for investigating excited state absorption, the methods could still be useful if analyzed carefully.
Especially, their predictions of ESA intensities should resemble those from multi-configurational methods given that the initial (here $S_1$) and final state of the process ($A_g$ in our case) are described by similar configurations.

To gain experimental insight into the orientation of the transition dipole moments for the two ESA transitions, and to deepen our understanding about the broadening of the second ESA peak with the potential shoulder at 4.93~eV, we determined the transient absorption anisotropy (TAA). 
We therefore measured the transient absorption for both parallel ($\Delta A_\parallel$) and perpendicular ($\Delta A_\perp$) pump and probe polarizations and calculated the TAA value as discussed in the Methods section \ref{sec:esa-graz} \cite{Lakowicz2006anisotropy,Zheng2020}. Note that erroneous values outside the range of -0.2--0.4 are possible in case of multiple contributing peaks of opposite sign \cite{Xu2023anisotropy}, as well as sample degradation.
This polarization-controlled TA measurement informs about the relative angle between the transition dipole moments of the pump and probe transitions, as detailed in the Methods Section.

Measured data for  $\Delta t = 0$~ps and $\Delta t = 10$~ps is presented in Fig.~\ref{fig:esa}c. Note that the value of 10~ps was chosen to account for shifts in the temporal pump-probe overlap $\Delta t_0$ caused by probe wavelength changes in the light source.

For an isotropic sample, following ref.~\cite{Lakowicz2006anisotropy,Xu2023anisotropy}, the anisotropy $r$ is expected to take a value of 0.4 (red dashed line in Fig.~\ref{fig:esa}c), if the transition dipole moment of the probed state is \textit{parallel} to the one of the pumped state.
In case the two dipole moments are \textit{perpendicular}, $r$ approaches $-0.2$  (purple dashed line in Fig.~\ref{fig:esa}c).
For the two main ESA features at 4.38~eV and 4.78~eV, we find a TAA signal around 0.4, indicating that both transitions are polarized along the same direction as the $S_1$, i.e. along the long molecular axis.
Thus, both peaks indeed belong to $A_g$ states.
For illustration, we have also modeled this behavior of $r$, shown as a dotted line in Fig.~\ref{fig:esa}c, using an artificial ESA spectrum of Cauchy-fitted peaks at the experimental energies. Nonphysical values outside the allowed range of \SIrange{-0.2}{0.4}{} are discussed in TAA method section.

In contrast, for the region beyond 4.8~eV, data from both measurement approaches show a trend towards lower TAA.
To explain this change, we have to include a third peak around 4.93~eV in our model, which is polarized orthogonal to the neighboring peak at 4.78~eV and has only one eighth of its intensity.
We admit that these scans still suffer from relatively high uncertainties exceeding $\Delta r>0.1$, reasons for which are discussed in TAA method section, owing to the noisy nature of the measured signals and possibly to some influence of a diffuse ground state depletion, which would be expected to increase with energy as can be seen in Fig.~\ref{fig:gsa}a, but did not result in observable peaks in this TA energy region.

We can think of two scenarios for the appearance of a high-energy shoulder in the second ESA peak.  
According to the first scenario, our TAA measurements reveal a change of polarization due to different electronic excitations. 
In this case, the rotation of the transition dipole from the long towards the short molecular axis would assign the shoulder to an excitation into a $B_{1g}$ state. 
However, as a counterargument to this interpretation, we were not able to locate an electronic state with $B_{1g}$ symmetry and comparable oscillator strength in this energy range in our calculations. 
In the second scenario, the shoulder in the ESA spectrum at 4.93~eV originates from a vibronic progression of its neighboring peak at 4.78~eV. 
This interpretation is supported by the energy difference of about 0.14~eV, which lies within the range expected for vibrational shoulders.
According to the TA data, the vibration then would have to have a non-parallel transition dipole compared to the lower electronic 4-$A_g$ state. 

\subsection{Two-Photon Absorption (2PA)}

Lastly, we want to complement our discussion with two-photon absorption (2PA) data, offers additional evidence for the assignment of the electronic states. 
Recalling the selection rules of two-photon absorption, transitions from the fully-symmetric ground state A$_g$ are allowed into other states with \textit{gerade} parity, that is, A$_g$ $\rightarrow$ A$_g$ and A$_g$ $\rightarrow$ B$_{g}$ \cite{ohira2008electronic}.
Thus, in the case of our squaraine, the 2PA spectrum should coincidentally feature transitions into the same states as the $S_1$ excited state absorption process.

In Fig.~\ref{fig:2PA}, The two-photon absorption cross section of SQIB in toluene and acetonitrile is plotted as a function of the absolute photon energy (in eV). These data show that the choice of solvent does not noticeably affect the 2PA response of SQIB, as the peak positions remain essentially identical in both solvents.
The measurement is based on two-photon excited fluorescence, i.e. the detection of light emitted after the simultaneous absorption of two photons. Although 2PA data for SQIB have been reported in the literature, the present measurements provide a more detailed view of the spectral response\cite{ohira2008electronic}.
Two weak peaks at 2.06 and 2.27~eV are attributed to vibronic transitions of the first excited state $S_1$.
Although direct 2PA into this B$_{3u}$ state is symmetry-forbidden, weak vibronically allowed transitions can occur through coupling with odd-parity vibrational modes of B$_u$ symmetry, which mix with the electronic B$_{3u}$ state and thus enable an effective A$_g$ $\rightarrow$ A$_g$ transition \cite{scherer2002two, ohira2008electronic}. 
The relative energetic distance of these features (0.21~eV) is reminiscent of the discussion of the ground-state absorption spectrum in Fig.~\ref{fig:S4}(a). 
A more intense peak is observed around 3~eV, which we assign to the 2-A$_g$ state. 
This assignment is consistent with the feature labeled as 2-A$_g$ at 3.15~eV in Region~II of the GSA spectrum discussed in Fig.~\ref{fig:gsa}(c).  
Finally, the highest measured point in the 2PA spectrum appears beyond 3.5~eV and has previously been estimated at 3.6~eV  \cite{AvilaFerrer2019}. 
Although the measurement could not be extended beyond this energy, it is reasonable to assume that the next transition would correspond to the 3-A$_g$ state, and thus, we would expect the maximum to appear actually around 4.38~eV in accordance with our ESA analysis.

\begin{figure}[ht]
    \centering
    \includegraphics[width=0.5\linewidth]{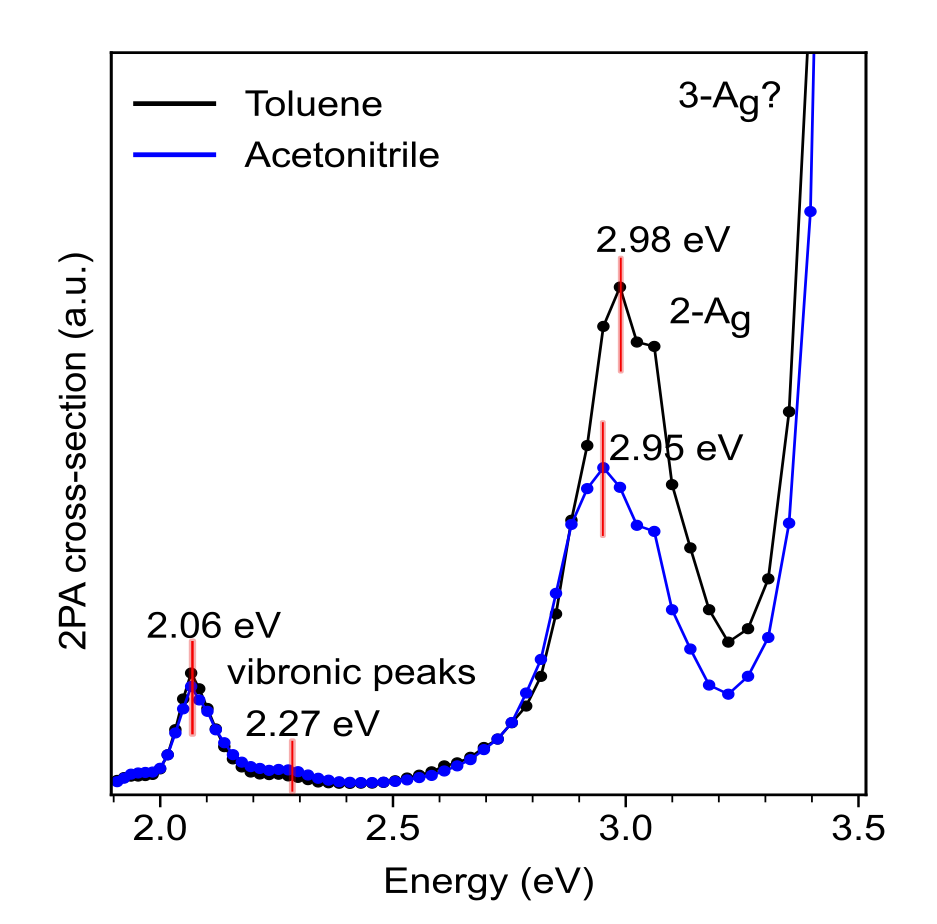}
    \caption{
Two-photon absorption (2PA) spectrum of SQIB in toluene and acetonitrile. 
The vibronic peaks of S$_1$ appear at 2.06~eV and 2.27~eV, followed by the main 2-A$_g$ 
transition at 2.98~eV in toluene and 2.95~eV in acetonitrile. 
The negligible energy shift between the two solvents indicates that the 2PA response of SQIB 
is largely insensitive to the solvent environment.
}
    \label{fig:2PA}
\end{figure}

\section{Conclusions}\label{sec3}

Squaraine chromophores are widely used in organic semiconductor materials not only for their intense visible--NIR absorption but also for their tunable optical response driven by controlled intermolecular coupling. Understanding the nature of their higher excited states and ESA channels is therefore essential for connecting molecular photophysics to macroscopic device performance. In this context, our work provides a comprehensive mapping of the optical excitation landscape of the isolated N-alkylated anilino squaraine (SQIB) by merging broadband spectroscopy with \textit{ab initio} methods.
Extending the conventional experimental window by additional 3~eV, we unveil previously unresolved high-energy transitions in ground state absorption.
Specifically, we identify multiple peaks at the onset of the UV-region around 3.5~eV. 
Besides electronic transitions, we link their origin to intensity enhancement of dipole-forbidden excitations due to coupling with vibrations. 
Moreover, we describe, for the first time, the rich spectral structure of squaraines beyond 5~eV.
We further elaborate on the energy position of the lowest three $A_g$ states of the molecule, which we locate between 3 and 5~eV, based on combined pump-probe and two-photon absorption data.

Finally, we note that the application of three substantially different theoretical methods (TD-DFT, GW/BSE and CASSCF/NEVPT2), a necessity arising from the challenging complexity of the electronic structure at hand, also allows for a direct comparison of their predictive power, accuracy and computational cost with respect to the photophysical properties of squarine-based dyes.

Despite the lack of a desired quantitative agreement, all three methods provide the same peak assignments for the experimentally observed ground-state absorption features. This agreement clearly corroborates our theoretical effort, and suggests the extension of our method comparison towards exited-state absorption. Performance on this task, however, is clearly compromised in the case of TD-DFT and GW/BSE due to their intrinsic limitation to singles excitations. The predictive power of TDDFT is additionally hampered by well-known resonance effects, which come to play in the case of an S$_1$ excited state absorption of similar energy as the excitation from the ground state into S$_1$. The spectra predictions of GW/BSE and CASSCF/NEVPT2, on the other hand, remain comparable in their deviation from the experiment, although their peak asignment differs, most likely due to the double-excitation character of A$_g$ states.

Covering such a broad spectral range, we hope that our description of the excitonic processes in squaraines forms a basis for future research into the photophysical properties of these molecules. The higher-lying states identified in this work will contribute to accurate modeling of processes in SQ materials, like energy- and charge transfer dynamics, exciton-exciton interactions, or singlet-singlet annihilation. \cite{Vlker2014exciton,Mal2023nonlinear}

\section{Method Section}\label{sec4}

\subsection{Experimental Details}

\subsubsection{Solution Sample and Transmission Spectra}
The anilino squaraine 2,4-Bis[4-(\emph{N,N}-diisobutylamino)-2,6-dihydroxy\-phenyl]squaraine (SQIB) was synthesized according to an established protocol\cite{Abdullaeva2016}. A stock solution of SQIB was prepared in amylene-stabilized chloroform at a concentration of 0.231 mM. From the stock solution 50 \textmu L or 500 \textmu L were transferred to glass vials for preparation of the "low concentration" and "high concentration" solutions, respectively. After the chloroform was removed under reduced pressure, 3000 \textmu L acetonitrile were added to the solid squaraine residue and stirred at 60 \textdegree C for 15 minutes. The solution was allowed to cool to room temperature and subsequently transferred to a \textit{Hellma} quartz glass cuvette with 1 cm optical path length. The resulting concentrations were 3.85 \textmu M (low concentration) and 38.5 \textmu M (high concentration) of SQIB in acetonitrile, respectively. The transmission spectra were recorded using a Specord 200 UV-Vis spectrometer (Analytic Jena) with baseline corrections against pure acetonitrile solvent. A measurement of pure acetonitrile against an empty cuvette showed an UV cut-off at around 6.2 eV (200 nm).

\subsubsection{PMMA-matrix Sample and Transmission Spectra}
“Solid solution” samples of SQIB were prepared in a PMMA matrix by spincoating on 1 mm thick transparent substrates in ambient conditions. For this poly-methyl-methacrylate (PMMA) with an average molecular weight of about 350,000 (g/mol) (Sigma Aldrich) was dissolved in chloroform to a concentration of 40 mg/mL. SQIB dissolved in amylene-stabilized chloroform was added to the PMMA solution to obtain concentrations of approximately 0.5, 1, and 4 percent by weight of the PMMA. The mixture was spincoated on float glass (VWR) and on fused silica (GE124 Ted Pella) substrates and subsequently dried at 60$^\circ\mathrm{C}$ on a hotplate. The transmission cut-off for the samples is approximately at 5.4 eV (230 nm) for fused silica and 3.65 (340 nm) for float glass substrates. The layer thicknesses were determined by variable angle spectroscopic ellipsometry (J.A. Woollam M2000DI, CompleteEASE software)\cite{Kamptner2024} to be approximately 6.5 eV (190 nm) and 1.90 eV (650 nm) on fused silica and on float glass, respectively. Normalized absorbance spectra (-log(T)) are shown in the Supporting Information in Fig.~\ref{fig:S6} together with a spectrum of SQIB dissolved in chloroform for reference. The spectrum in chloroform is typical of diluted monomeric SQIB in solution showing a strong $S_0 \rightarrow S_1$ transition and including a vibronic progression shoulders. When SQIB is embedded in a PMMA matrix the spectral profile broadens and the vibronic progression gains intensity as the weight percentage of SQIB increases as a consequence of a reduced average distance of the molecules\cite{Zheng2020}.

\subsubsection{ESA Spectra}
\label{sec:esa-graz}

Transient absorption spectra were recorded using an amplified laser system (Light Conversion PHAROS PH1-20, \SI{1030}{nm} central wavelength, \SI{40}{\kilo\hertz} pulse repetition rate, \SI{400}{\micro\joule} pulse energy), powering two non-collinear optical parametric amplifiers, one for the pump  beam (Light Conversion ORPHEUS-N-2H) and one for the probe beam (ORPHEUS-N-3H). A detailed description of the setup is given in ref.~\cite{schwarzl2022transient}. Probe wavelengths from \SIrange{250}{470}{\nano\meter} (\SIrange{4.96}{2.64}{\electronvolt}) were generated by frequency doubling the ORPHEUS-3H output in a \SI{0.5}{\milli\meter} BBO crystal (EKSMA BBO-603H or BBO-611H), the fundamental wavelength was discarded using 3 mm Schott glass filters (UG5, BG39 and BG3). UV fused silica lenses ($f=\SI{125}{\milli\meter}$) were used for coarse focusing and chromatic aberration was corrected by monitoring the focal plane with a UV camera (Artray ARTCAM-092UV-WOM).

No excitation of the PMMA matrix or on the glass substrate by the pump pulses is expected due to a lack of excited states in the visible range and the relatively low pump intensity insufficient for multi-photon excitation (\SIrange{2}{30}{\micro\joule\per\square\centi\meter}). Measurements were normalized to equal pump fluences and the ground-state depletion shoulder at a probe photon energy of 1.82 eV. Repeated delay scans were used to quantify sample deterioration and new spots were chosen frequently.

The transient absorbance as function of the delay time between pump and probe pulses, $\Delta A (\Delta t)$, for one combination of pump and probe wavelengths is modeled by the product of a Gaussian cumulative distribution function (normcdf) accounting for the instrument response, and a sum of multiple exponential decays:
\begin{equation}\label{eq:ta-exp-model}
    \Delta A (\Delta t) = \text{normcdf} (\Delta t, \Delta t_0, \sigma_t)*\sum_{i=1}^n A_i \exp \left( -\frac{\Delta t-\Delta t_0}{\tau_i} \right)
\end{equation}
We fit the temporal overlap position $\Delta t_0$, the rise time $\sigma_t$, and the 1-3 exponential decays consisting of amplitude $A_i$ and decay time $\tau_i$ to the experimental results. The total amplitude, neglecting the constant factor 0.5 given by the instrument response function, can then be calculated as $\Delta A_\text{tot}=\sum_i A_i$, or the fit function can be evaluated at any desired delay time. Owing to the spectral characteristics of the Orpheus-2H pump light source with a peak width of 75 meV, there is a spectral overlap of 12\% between the 653~nm ($\sim$1.90 eV) pump and the 0-1 vibronic transition of 1\% SQIB in PMMA. Theoretically, this could result in small red-shifted ESA peak shoulders, which were not visible in the measurements in Fig.~\ref{fig:esa}a.

\subsubsection{Transient Absorption Anisotropy (TAA)}
\label{sec:TAA}
In isotropic samples investigated using transient absorption, the relative linear polarization of pump and probe pulses introduces an effect known as \emph{transient absorption anisotropy}. Similar to fluorescence anisotropy~\cite{Lakowicz2006anisotropy}, it provides information about the relative orientation of the transition dipole moments of pumped and probed states, as well as randomization through rotation of molecules in solution, or excitation transfer between neighboring chromophores in solid matrices.\cite{Zheng2020} A recent investigation by Xu et al.\cite{Xu2023anisotropy} shows that TAA has more edge cases, such as when superimposed positive and negative TA signals result in diverging TAA signals. The publication provides a standard definition of the time- and photon energy dependent TAA $r(\Delta{}t, E_\text{pump}, E_\text{probe})$, where we adapted the arbitrary signals $I$ to match the transient absorbance in the parallel $\Delta A_\parallel$, and the perpendicular case $\Delta A_\perp$, like in ref.~\cite{Zheng2020}:
\begin{equation}
    r (\Delta{}t, E_\text{pump}, E_\text{probe}) = \frac{\Delta A_\parallel - \Delta A_\perp}{\Delta A_\parallel + 2\Delta A_\perp} = \frac{2}{5} \cdot \frac{3 \cos^2 \theta - 1}{2}
\end{equation}
This definition provides a relation between the relative angle $\theta$ of the transition dipole moments of pumped and probed states, averaged over the investigated ensemble. The low signal-to-noise ratio of ESA states measured in SQIB in PMMA, combined with the possibility of sample degradation and diffuse contributions from the ground state bleach\cite{Xu2023anisotropy}, is also expected to contribute to nonphysical results outside the valid range from -0.2 (completely perpendicular, $90^\circ$), via 0.0 (random or contributions adding up to the magic angle) to 0.4 (completely parallel, $0^\circ$).

Measurements were performed by averaging over many scans, removing residual fluorescence and transmitted pump signal by subtracting the probe-pump signal to the left of the temporal overlap. For selected probe photon energies, we performed delay scans and fitted them with a bi-exponential function according to eq.~\ref{eq:ta-exp-model}. For these scans, the transient anisotropy was calculated from the fit result. Other measurements were recorded at one fixed delay before and after the temporal overlap to gain more statistics. Due to the NOPA light sources' characteristics, a shift of the temporal overlap by a few ps is expected when changing the probe wavelength which could not be accounted for in these fixed point measurements, resulting in a varying amount of randomization through FRET.

\subsubsection{Two-Photon Absorption (2PA)}
The two-photon absorption spectrum of SQIB in diluted toluene and acetonitrile solution (~\SI{6e-7}{M}) 
was collected adopting the two-photon excited fluorescence (2PEF) technique. The measurement was carried out using a Nikon A1R MP+ multiphoton upright microscope and a tunable (700-1300 nm) femtosecond mode-locked laser (Coherent Chameleon Discovery) as the excitation source. The excitation beam was focused on the sample, contained in a quartz fluorescence cuvette, through a 25x (NA = 1.1) water dipping objective. The outcoming 2PEF signal was collected in epifluorescence mode. The quadratic dependence of the two-photon absorption process on the excitation power was verified over all the excitation range.

\subsection{Computational Details}

All electronic-structure calculations were performed using a hierarchical approach that combines density-functional, many-body perturbation, and multireference wavefunction methods. 
First, the geometries of both the ground (S$_0$) and first excited (S$_1$) states were optimized using the CAM-B3LYP functional \cite{yanai2004new} and the cc-pVTZ basis set \cite{dunning1989gaussian} within the ORCA~5.0.4 package~\cite{ORCA, neese2022software}. 
This hybrid functional and basis combination provides a reliable balance between accuracy and computational efficiency for $\pi$-conjugated chromophores~\cite{consiglio2024optimizing,jacquemin2008td}. 
To reduce computational cost, the isobutyl substituent of SQIB was replaced by a methyl group; as shown in Fig.~\ref{fig:S2}, this substitution has a negligible influence on the electronic structure and optical spectra. 

The ground-state absorption (GSA) spectrum was computed using linear-response time-dependent density functional theory (TDDFT) in ORCA, while the excited-state absorption (ESA) spectrum was obtained from quadratic-response TDDFT (QR-TDDFT) calculations performed with DALTON~2020~\cite{aidas2014d}. Due to the substantially higher computational cost and runtime of QR-TDDFT calculations in DALTON, the 6-311G$^\ast$ \cite{krishnan1980self} basis set was employed for the ESA calculations.
The QR-TDDFT formalism enables direct evaluation of  excited-state transitions and nonlinear optical responses, providing access to ESA  that are inaccessible within standard linear-response TDDFT~\cite{bowman2017excited}. 

To include many-body effects and improve the description of quasiparticle and excitonic states, calculations were performed within the self-consistent GW framework followed by the Bethe--Salpeter equation (BSE), as implemented in the Fiesta package~\cite{blase2011first,blase2011charge}. 
The same CAM-B3LYP/cc-pVTZ reference was used as the DFT starting point to ensure internal consistency across all theoretical levels. 
The GW/BSE formalism, which explicitly accounts for screened electron--hole interactions, has been widely benchmarked for optical excitations in molecular systems~\cite{jacquemin2015benchmarking}. 
To obtain the ESA spectra, we have implemented a method similar to the pseudo-wavefunction formulation of LR-TDDFT.
In BSE, the wave-function of an excited state, $\psi_{m}$, is usually described as a linear combination of electron-hole pairs

\begin{equation}
    \psi^{(m)}(r_h,r_e) = \sum_{v,c}[X_{vc}^{(m)}\phi_{v}^{\ast}(r_h)\phi_{c}(r_e) + Y_{vc}^{(m)}\phi_{v}^{\ast}(r_e)\phi_{c}(r_h)]
\end{equation}

Here, $\phi$ goes over the $v$ occupied and $c$ unoccupied orbitals of the system weighted by the eigenvector coefficients $X_{vc}$ and $Y_{vc}$ of the BSE solution for the $m^{th}$ excited state.
In this ansatz, the transition dipole, $\mu_{m,n}$, between two excited states $m$ and $n$ can be written as~\cite{Nascimento2022computational, Segatta2021insilico}

\begin{equation}
    \mu_{m,n} = \sum_{vc}\sum_{v'c'} [(X_{vc}^{(m)}X_{v'c'}^{(n)} - Y_{vc}^{(m)}Y_{v'c'}^{(n)}) (M_{cc'} \delta_{vv'} - M_{vv'} \delta_{cc'})]
\end{equation}

where $M$ represents the dipole operator acting on the molecular orbital basis and $\delta$ is the Kronecker-delta.

For a multireference description of both GSA and ESA, complete active space self-consistent field (CASSCF) and $n$-electron valence second-order perturbation theory (NEVPT2) calculations were carried out using MOLPRO~2024.1~\cite{werner2020molpro}. 
In the CASSCF method,\cite{werner80,werner81,werner85,knowles85} static electron correlation is treated explicitly by optimizing both configuration-interaction coefficients and molecular orbitals within a defined active space. 
Due to computational constraints, the full $\pi$-manifold could not be included; instead, a reduced active space of twelve electrons in twelve orbitals [CAS(12,12)] was used to make the calculations feasible. 
The molecular orbitals selected for the active space are illustrated in Fig.~\ref{fig:S1} and were chosen based on their $\pi$ and $\pi^*$ character, which dominate the low-lying excited states of SQIB. 
Following the CASSCF calculations, NEVPT2~\cite{angeli2001introduction,schapiro2013assessment} was applied to account for dynamic electron correlation beyond the CASSCF reference. 
It should be emphasized that careful selection of molecular orbitals and appropriate sizing of the active space are essential for obtaining accurate multiconfigurational results, particularly for the balanced description of both ground and excited states. 

\section{Author contributions}
NT, PP, MK and AW conceived the research. NT and AW performed all calculations and analyzed them with contributons from AWH and PP. MFS synthesized the SQIB compound and recorded the solution ground state spectra under the supervision of AL. MS prepared the PMMA-matrix samples and recorded the ground state transmission spectra and contributed to the manuscript writing.  RS, FL, MJ and MK performed transient absorption measurements. BB performed 2PA measurements. NT and AW wrote the manuscript with important contributions from all authors. 
\section{Note}
The authors declare no competing financial interest.

\begin{acknowledgement}
The work is funded by the Austrian Science Fund via FWF Grants 10.55776/PAT1379523 and 10.55776/P36903. MS thanks the Linz Institute of Technology (LIT-2023-12-SEE-114 SLIME) for funding. MFS and AL are obliged to the DFG GRK 2591 “Template-designed Organic Electronics – TIDE” for financial support. MFS thanks the Manchot Foundation for a doctoral scholarship. AWH is grateful to NAWI Graz for support and acknowledges the use of HPC resources provided by the IT services of Graz University of Technology (ZID).
BB acknowledges the equipment and framework of the COMP-R initiative, funded by the ''Departments of Excellence program of the Italian Ministry for University and Research'' (MUR, 2023-2027). We acknowledge ELI-ERIC for experimental and financial support during measurements at the ELI Beamlines under the projects ELIUPM4, TRANSDF and SQUAREPOP, especially Shirly Josefina Espinoza Herrera, Saul Vazquez Miranda and Mateusz Rebarz at the trELIps setup. We thank Anna Painelli for carefully reading the manuscript and providing helpful comments.
\end{acknowledgement}


\clearpage
\appendix
\section*{Supporting Information}
\setcounter{figure}{0}
\setcounter{table}{0}
\renewcommand{\thefigure}{S\arabic{figure}}
\renewcommand{\thetable}{S\arabic{table}}



\begin{figure}[h!]
\centering
\includegraphics[width=0.5\linewidth]{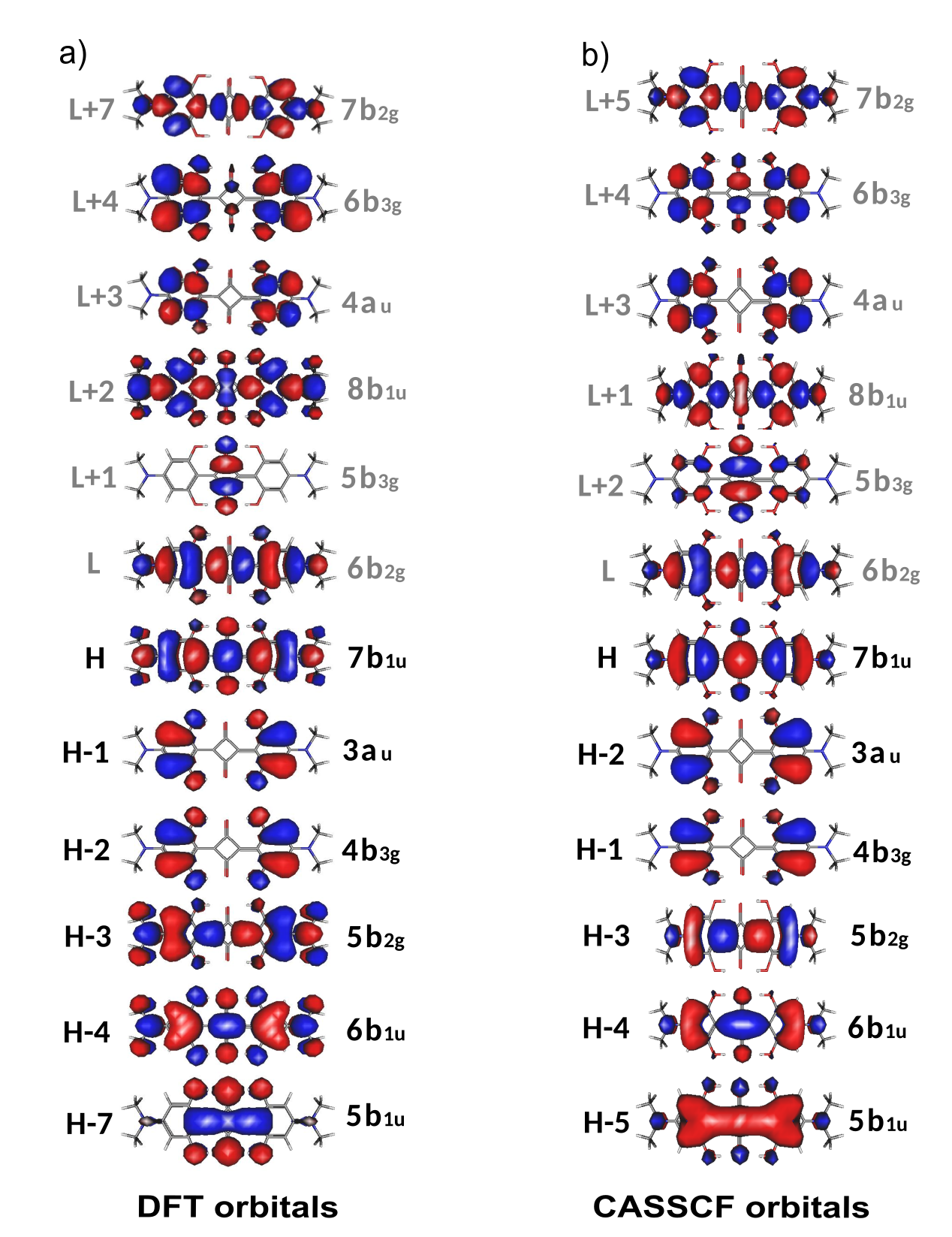}
\caption{
Comparison of selected molecular orbitals of the MeSQ monomer calculated using (a) DFT (CAM-B3LYP/cc-pVTZ) and (b) CASSCF(12,12)/cc-pVTZ. The orbital isosurfaces are plotted with an isovalue of 0.02 . Orbitals are labeled relative to the HOMO (H) and LUMO (L), together with their irreducible representations under $D_{2h}$ symmetry. Both methods yield the same set of orbitals with overall similar character, but with slight differences in their energetic ordering, leading to changes in HOMO and LUMO indexing. 
}
\label{fig:S1}
\end{figure}


\begin{figure}[h!]
\centering
\includegraphics[width=0.9\linewidth]{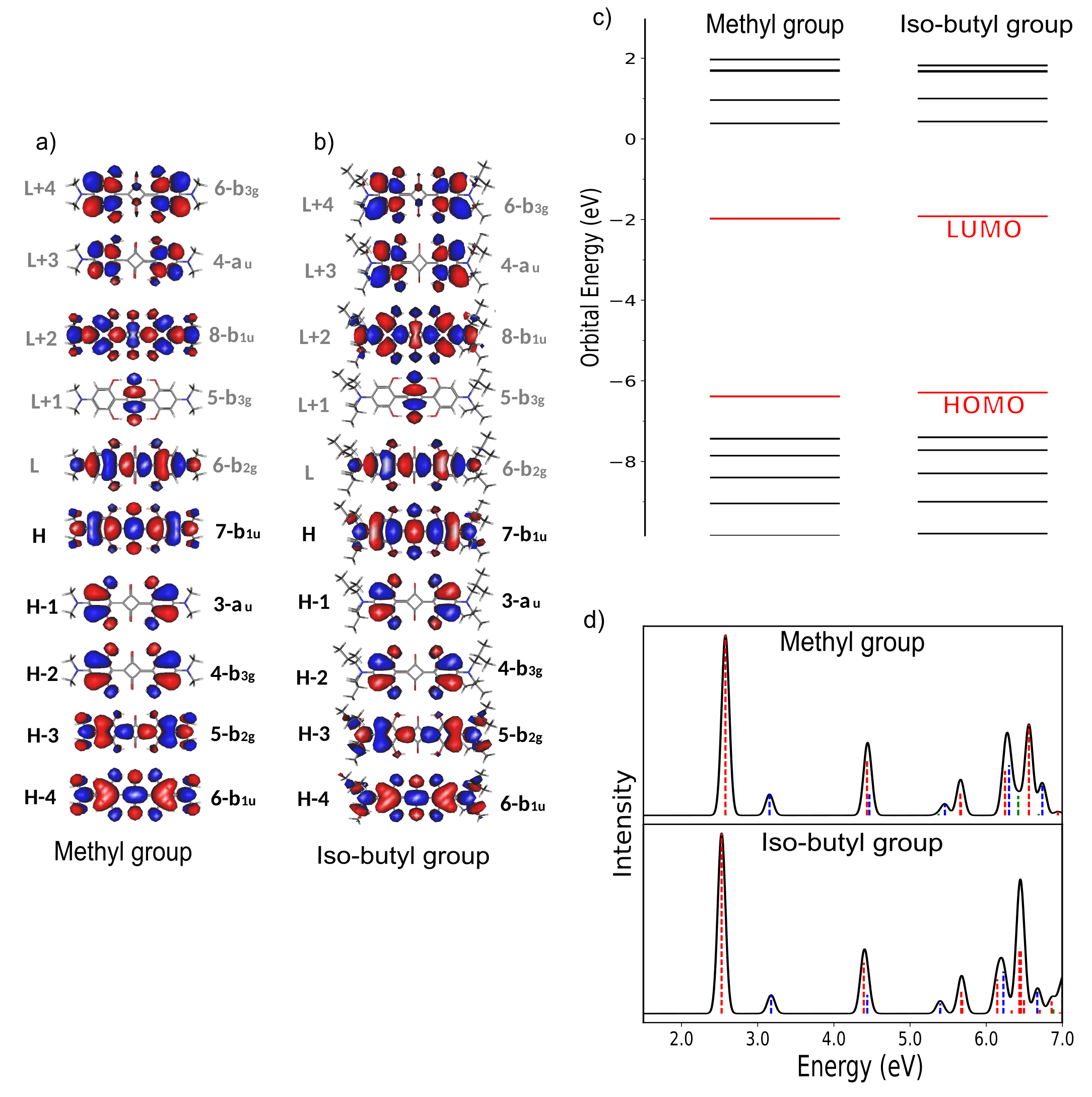}
\caption{
Comparison of the electronic structure and ground-state absorption (GSA) spectrum of the SQ molecule functionalized with methyl and iso-butyl groups. Panels (a) and (b) show the molecular orbitals from HOMO$-4$ to LUMO$+4$ for the methyl and iso-butyl derivatives, respectively, calculated using DFT (CAM-B3LYP/cc-pVTZ). The orbitals exhibit identical symmetry labels and similar shapes, with an isosurface value of 0.02, indicating that the functional group has minimal impact on the frontier orbitals. Panel (c) presents the orbital energy diagrams, where HOMO and LUMO levels are highlighted in red. The orbital energies are nearly the same for both systems. Panel (d) compares the normalized ground-state absorption (GSA) spectra of both molecules, confirming that their spectral features are essentially identical and only minor deviations appear above 6~eV.
}
\label{fig:S2}
\end{figure}


\begin{figure}[h!]
\centering
\includegraphics[width=0.9\linewidth]{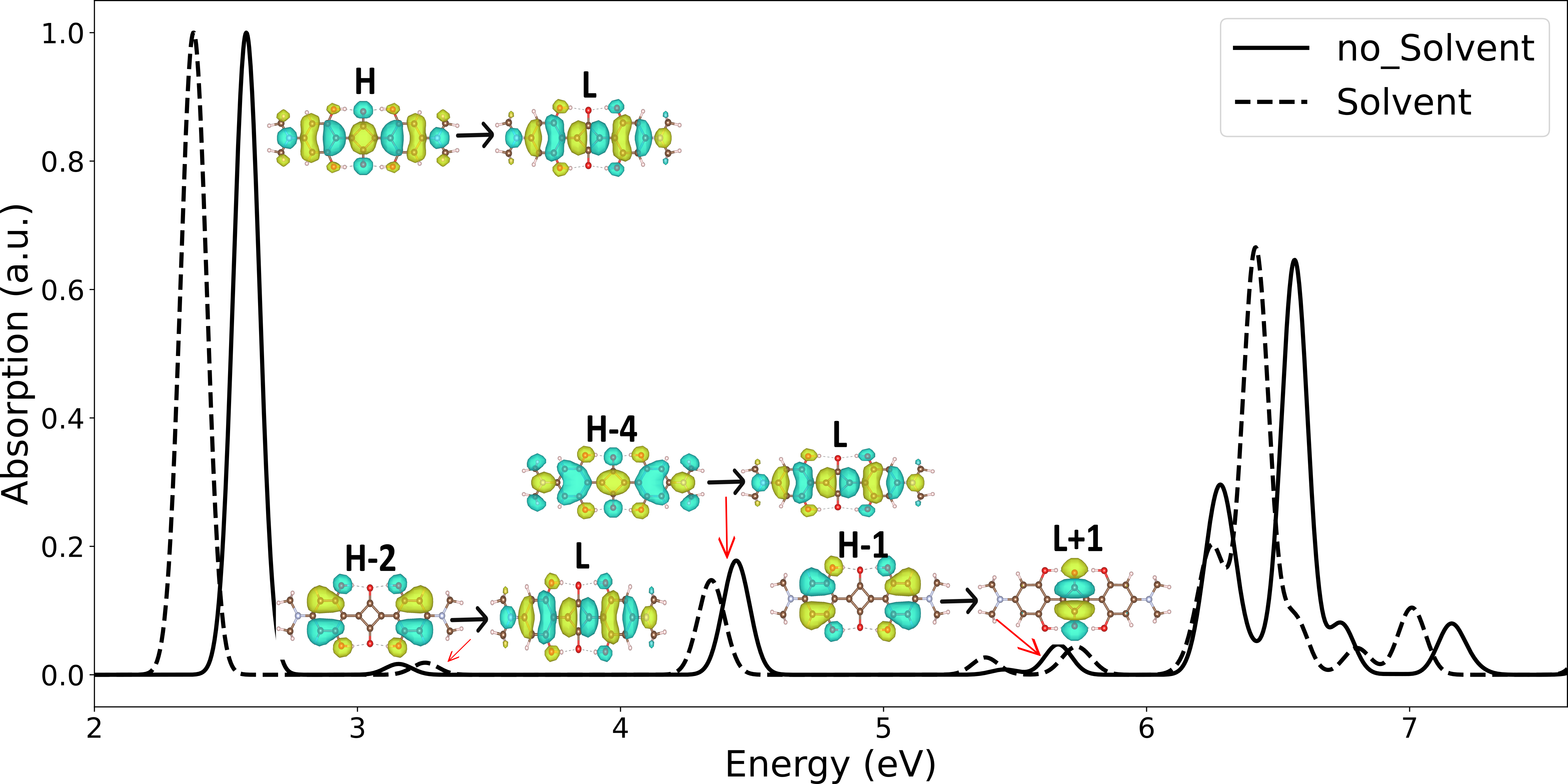}
\caption{Absorption spectra calculated by TDDFT (ORCA, CAM-B3LYP/cc-pVTZ, conductor-like polarizable continuum model (CPCM)) without solvent (solid line) and with solvent (dashed line). The plotted orbital isosurfaces correspond to representative transitions and illustrate the electronic character of the excitations.The effect of solvent on the absorption spectrum is not a simple uniform shift but depends on how the solvent interacts with the electronic density redistribution of each excitation. In the CPCM, the magnitude and direction of the shift are determined by the orbital character of the transition. For the first intense peak (HOMO$\rightarrow$LUMO, $\pi \rightarrow \pi^*$), the delocalized $\pi$ and $\pi^*$ orbitals interact strongly with the reaction field, which lowers the excitation energy and produces a red shift. In contrast, the second peak involves more localized orbitals that couple less efficiently with the solvent polarization; here the excitation energy increases, giving a blue shift. Thus, the same solvent can induce different spectral responses depending on the electronic nature of the transition.}
\label{fig:S3}
\end{figure}

\FloatBarrier           


\subsection*{Vibronic spectra}
The vibrationally resolved GSA spectra for Regions~I and~II (Figures.~\ref{fig:S4}, \ref{fig:S5}) were computed with the \textit{Excited-State Dynamics} (ESD) module of \textsc{ORCA}~5.0 within the Franck--Condon approach. 
The Adiabatic Hessian After a Step (AHAS) approach was employed,
which uses the ground-state (GS) geometry and Hessian to approximate the excited-state (ES)
Hessian after optimization step. All calculations were carried out
at the  CAM-B3LYP/6-311G* level of theory, and Herzberg--Teller (HT) terms were included. The spectra and rate constants were obtained from
Fermi’s Golden Rule using the path-integration formalism (derivations in Ref.\cite{de2019predicting}). All vibronic simulations were performed in the zero-temperature. We considered Duschinsky rotation (mode mixing) for considering differences between ground- and excited-state normal-mode frequencies.
The displacement vector and Duschinsky rotation matrix are defined as
\[
K = L_x^{\mathsf T}(\bar{q}_0 - q_0), \qquad
J = L_x^{\mathsf T}\bar{L}_x ,
\]
where $L_x$ and $\bar{L}_x$ are the eigenvectors of the mass-weighted Hessians at the GS and ES
geometries, respectively, and $q_0$ and $\bar{q}_0$ are the corresponding mass-weighted
Cartesian nuclear coordinates.

In Figure~\ref{fig:S4}(a), the experimental spectrum is compared with the vibronic profile simulated using
the AHAS method in region I. The calculated vibronic progression is assigned by its dominant vibrational
modes. The Huang--Rhys factors (HRFs), obtained from the ESD module, identify modes 20, 44,
55, 73, and 123 as the primary contributors, with modes 20 and 55 showing the strongest
couplings. 

In Figure~\ref{fig:S4}(b), the dimensionless displacements are plotted against vibrational 
wavenumber (cm$^{-1}$). The largest displacements are found for mode~20 (182~cm$^{-1}$) 
with $\Delta = 0.52$ and mode~55 (670~cm$^{-1}$) with $\Delta = 0.19$, in good 
agreement with the experimental bands reported at 147 and 570~cm$^{-1}$
and at 166 and 574~cm$^{-1}$~\cite{burigana2025signatures,timmer2022charge}. The slightly higher frequencies obtained in our calculations are mainly related to the different terminal substituents attached to the squaraine core. In the present system, the light methyl groups reduce the effective reduced mass of the vibrational modes, which directly leads to higher wavenumbers according to the inverse mass dependence of vibrational frequency. In contrast, the molecules studied in previous works contain heavier and more complex substituents, resulting in larger reduced masses and therefore lower vibrational frequencies for the corresponding modes. The corresponding normal 
modes are highlighted and visualized on the right: the mode at 182~cm$^{-1}$ corresponds 
to a symmetric stretching motion of the two arms relative to the squaric acid center, 
while the strong mode at 670~cm$^{-1}$ represents a stretching vibration that distorts 
the central benzene rings.

In Figure~\ref{fig:S5}, the vibrationally resolved spectrum of Region~II is presented. 
The experimental spectrum in panel~(a) exhibits four distinct peaks. 
As discussed in the main text, peaks~3 and~4 are assigned to the electronic 
states 2-B$_{3u}$ (red) and 2-B$_{2u}$ (blue), respectively. 
In contrast, the grey peaks~1 and~2 cannot be attributed to purely electronic 
excitations. Instead, these features are assigned to Herzberg--Teller (HT) activity 
of the 1-B$_{3g}$ and 2-A$_g$ states, which are symmetry-forbidden within the 
Franck--Condon (FC) approximation. 

This assignment is supported by panels~(b) and~(c), where the simulated vibronic 
spectra of the 1-B$_{3g}$ and 2-A$_g$ states show intensity arising exclusively 
from HT contributions. Because the Herzberg--Teller intensities are generally much 
weaker than the Franck--Condon ones, the spectra in panels~(b) and~(c) were multiplied 
by a factor of twenty and ten respectively , to allow a direct comparison with the FC-dominated spectra in 
panels~(d) and~(e). Panels~(d) and~(e) show the vibronic spectra of the 2-B$_{3u}$ 
and 2-B$_{2u}$ states, respectively, which are dominated by Franck--Condon activity, 
consistent with their symmetry-allowed character. 

The calculated profiles further reproduce the experimental broadening trends: 
the broad feature associated with peak~4 agrees well with the simulated 
2-B$_{2u}$ spectrum in panel~(e), while peak~3 also exhibits less broadening 
in both experiment and theory. This agreement supports the assignment of peaks~3 
and~4 to the 2-B$_{3u}$ and 2-B$_{2u}$ states, respectively. However, the situation 
for peaks~1 and~2 is less straightforward. All calculated vibronic bands in this energy 
range exhibit significant overlap, and the HT-derived intensities are roughly an order 
of magnitude smaller than the FC ones. In the experimental spectrum, peak~2 displays an 
intensity comparable to peaks~3 and~4, suggesting an alternative interpretation. 
In this scenario, peak~2 may represent a vibronic shoulder of the 2-B$_{3u}$ transition, 
arising from multiple vibrational mode contributions rather than a separate HT-induced 
transition. The very weak peak~1, on the other hand, can still be reasonably assigned to 
minor HT activity involving the 1-B$_{3g}$ and 2-A$_g$ states, as indicated by the broad 
features in panels~(b) and~(c).


\begin{figure}[h!]
\centering
\includegraphics[width=0.5\linewidth]{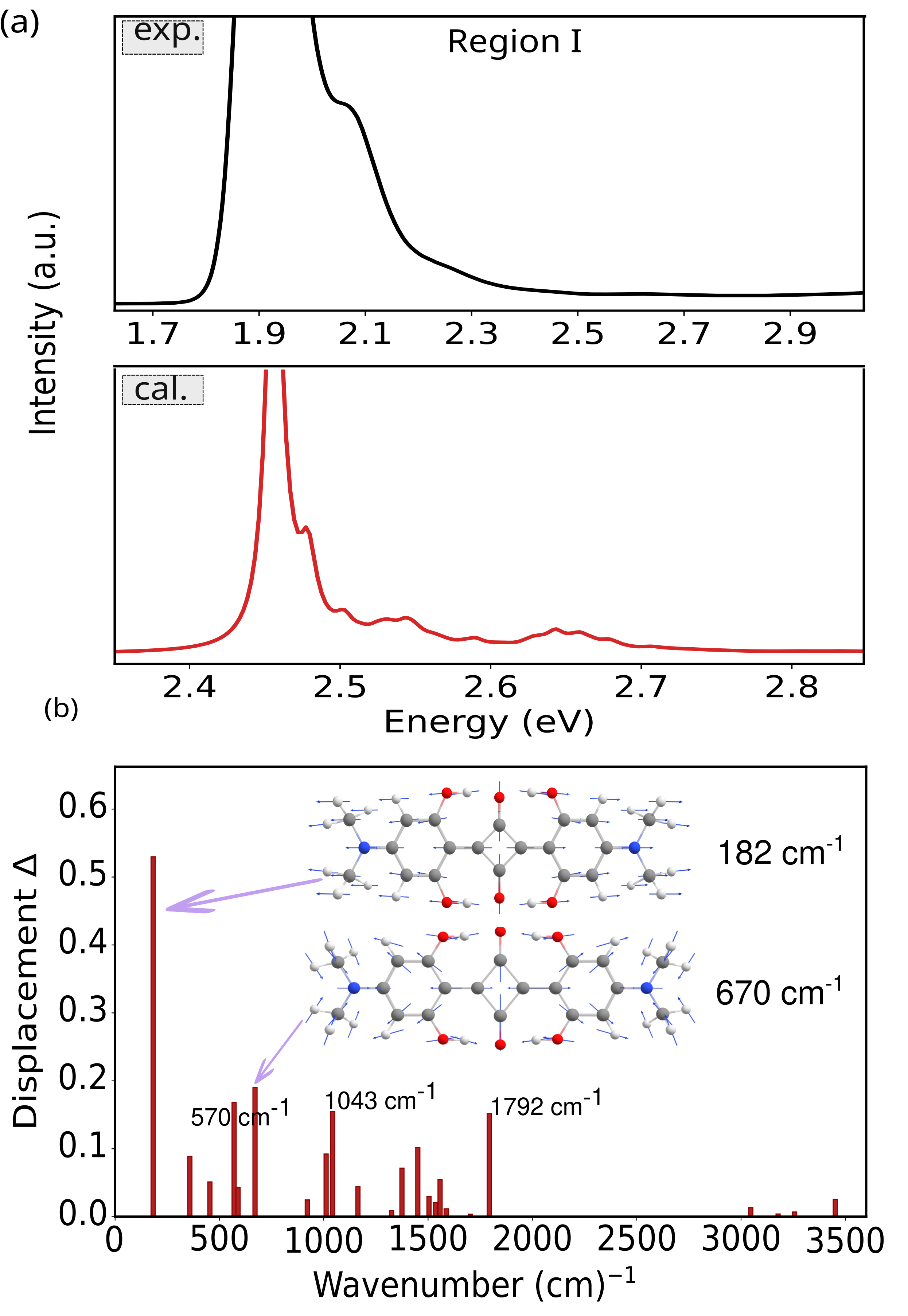}
\caption{(a) Experimental GSA spectrum compared with the vibronic profile simulated 
using the AHAS method (CAM-B3LYP/6-311G* with HT terms). (b) Dimensionless displacements vs.\ wavenumber, with 
the vibrational mode~20 (182~cm$^{-1}$, $\Delta=0.52$) and mode~55 (670~cm$^{-1}$, $\Delta=0.19$) 
depicted in the insets.}
\label{fig:S4}
\end{figure}


\begin{figure}[h!]
\centering
\includegraphics[width=0.6\linewidth]{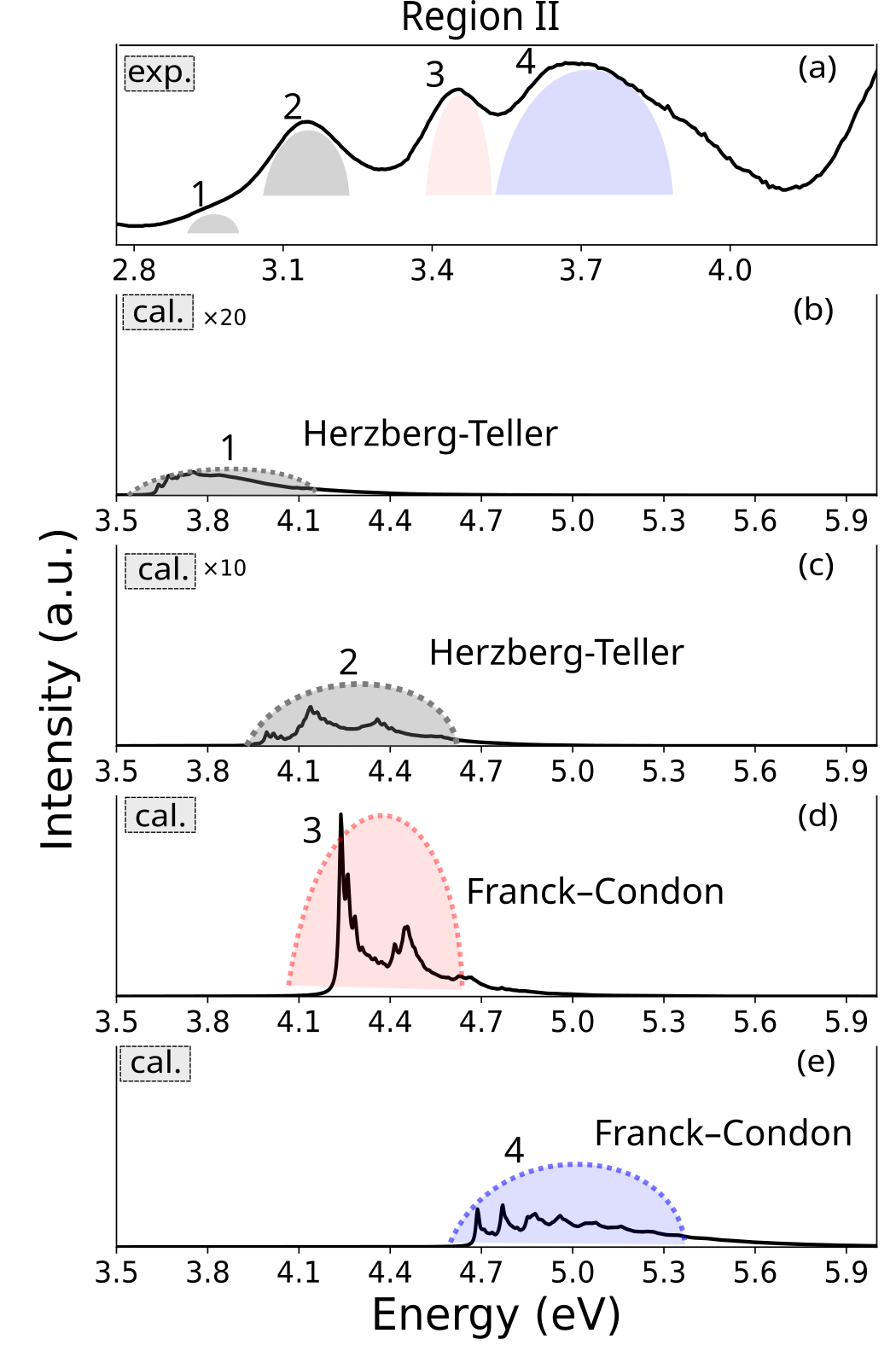}
\caption{Vibrationally resolved GSA spectrum of Region~II. 
(a) Experimental spectrum showing four distinct peaks. 
Peaks~3 and~4 are assigned to the 2-B$_{3u}$ (red) and 2-B$_{2u}$ (blue) 
electronic states, respectively, while the grey features (peaks~1 and~2) 
originate from Herzberg--Teller effect associated with the 1-B$_{3g}$ and 2-A$_g$ states. 
(b,c) Simulated vibronic spectra of the 1-B$_{3g}$ and 2-A$_g$ states, 
displaying intensity solely from Herzberg--Teller contributions. 
(d,e) Simulated vibronic spectra of the 2-B$_{3u}$ and 2-B$_{2u}$ states, 
dominated by Franck--Condon in line with their symmetry-allowed character.}
\label{fig:S5}
\end{figure}

\clearpage


\begin{table}[htbp]
\centering
\caption{List of all symmetry-allowed electronic transitions that are not included in the main text. The corresponding oscillator strengths below 0.01 were not reported. }
\label{tab:S1}
\begin{tabular}{lccc}
\toprule
\textbf{Symmetry} & \textbf{TDDFT (eV)} & \textbf{GW (eV)} & \textbf{CASSCF (eV)} \\
\midrule
1--B$_{2u}$ & 3.16 & 2.88  & 2.79 \\
3--B$_{2u}$ & 5.46 & 5.26  & 5.01 \\
5--B$_{3u}$ & 6.25 & 6.10  & 5.37 \\
2--B$_{1u}$ & 6.42 & 6.505 & --   \\
5--B$_{2u}$ & 6.74 & 6.502 & 6.12 \\
6--B$_{3u}$ & 6.94 & 6.52  & 5.86 \\
7--B$_{3u}$ & 7.15 & 6.94  & 5.89 \\
6--B$_{2u}$ & 7.16 & 6.904 & 6.38 \\
7--B$_{2u}$ & 7.73 & 6.84  & 6.67 \\
8--B$_{2u}$ & 7.91 & 7.62  & 6.69 \\
8--B$_{3u}$ & 7.92 & 7.54  & 6.21 \\
9--B$_{3u}$ & 8.35 & 8.00  & 6.39 \\
\bottomrule
\end{tabular}
\label{table:S1}
\end{table}
\clearpage


\subsection{Absorbance Spectra of SQIB in Solution and PMMA Films}

\begin{figure}[h!]
\centering
\includegraphics[width=0.8\textwidth]{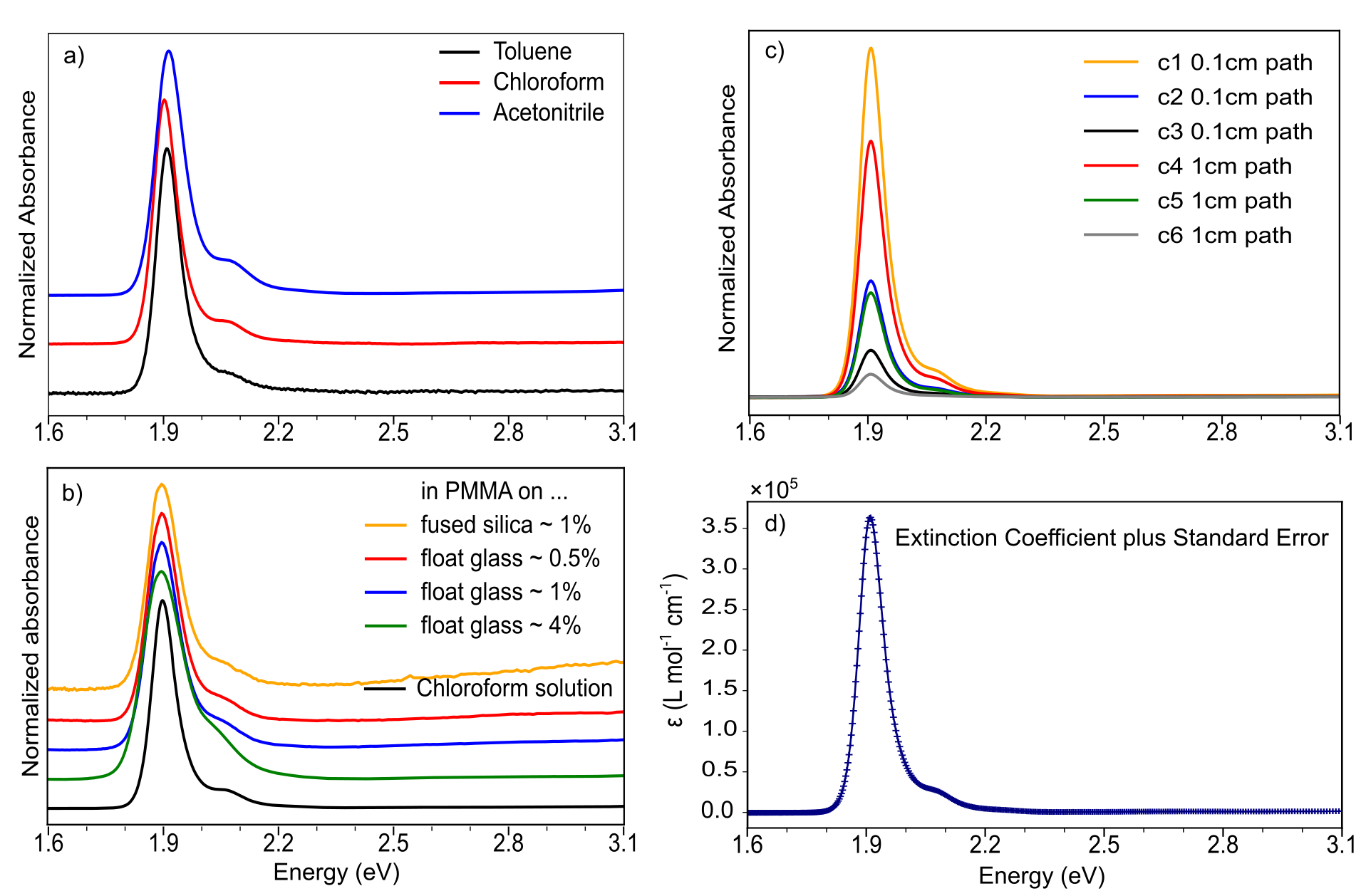}
\caption{
(a) Normalized absorbance spectra of SQIB in different solvents (toluene, chloroform, and acetonitrile). 
(b) Normalized absorbance spectra of SQIB embedded in a PMMA matrix on different substrates, together with the spectrum in chloroform solution. 
(c) Six transmission spectra ($T$) were recorded at different concentrations: 
$c_1 = 4.2 \times 10^{-5}$ mol L$^{-1}$, 
$c_2 = 1.38 \times 10^{-5}$ mol L$^{-1}$, 
$c_3 = 5.47 \times 10^{-6}$ mol L$^{-1}$, 
$c_4 = 3.05 \times 10^{-6}$ mol L$^{-1}$, 
$c_5 = 1.23 \times 10^{-6}$ mol L$^{-1}$, and 
$c_6 = 2.57 \times 10^{-7}$ mol L$^{-1}$. 
The first three spectra ($c_1$--$c_3$) were measured using a 1 mm cuvette, while the remaining three ($c_4$--$c_6$) were acquired with a 1 cm cuvette. 
All spectra were referenced against a toluene-filled cuvette. 
The absorbance $A = -\log(T)$ at each energy was analyzed according to Lambert--Beer's law \cite{beer1852bestimmung}, $A = \varepsilon \, c \, l +b(E)$ where $b(E)$ accounts for an energy-dependent baseline offset. (d) To obtain $\varepsilon(E)$ without performing individual fits at each energy value, a global linear least-squares procedure was applied: for every energy point, the vector of absorbance values was regressed against the product $c \, l$ using the Moore--Penrose pseudoinverse\cite{penrose1955generalized}. 
This yields the spectrum of the molar extinction coefficient $\varepsilon(E)$ directly from all six measurements simultaneously. 
The standard error of $\varepsilon(E)$ at each energy was calculated from the residual variance of the regression, providing a quantitative estimate of the uncertainty of the fitted extinction coefficient.}
\label{fig:S6}
\end{figure}
\clearpage

\subsection*{White light-based TA spectrum}
For comparison to the narrow-band measurements of 1\% SQIB in PMMA thin films, white light measurements of 0.5\% SQIB in PMMA were recorded at the trELIps setup \cite{Richter2021trelips} at ELI Beamlines. A white light (WL) supercontinuum was generated in a CaF$_2$ crystal, which was used as probe pulse. The TA spectrum was dispersed in prisms and recorded using a set of laser-synchronized Stresing high-speed cameras. Pixel-to-wavelength calibration was performed by fitting Gaussian spectral distributions to the transmission of 18 band-pass filters, followed by a polynomial fit to interpolate all other pixels.

Figure~\ref{fig:S7} shows the resulting TA measurement on 0.5~\% SQIB in PMMA on glass, averaged over the first ps of delay times, upon excitation by an OPA at 650~nm at approx. \SI{200}{\micro\joule\per\square\centi\meter}. Visible features were fitted with six normalized Gaussian PDF functions. Note that the small 1.5~eV feature shows wide variations between measurements, indicating its origin in instabilities in the the WL region, and was therefore excluded using fit boundaries. The S1 excitation at 1.92~eV and its vibronic shoulder at 1.98~eV are clearly visible. At 2.59~eV and 2.90~eV, corresponding to total energies of 4.49 and 4.80~eV, two ESA features can be seen, matching the narrow-band-based results visible in Fig.~4 within uncertainty. However, the shoulder of the higher-energy ESA peak is placed outside the measurement range by the fit algorithm. The fit start parameters and bounds were tuned to avoid fitting a diffuse, broad background spanning both ESA peaks.  In the final fit result, no parameter reached the limits of the boundaries set.

\begin{figure}
    \centering
    \includegraphics[width=0.9\linewidth]{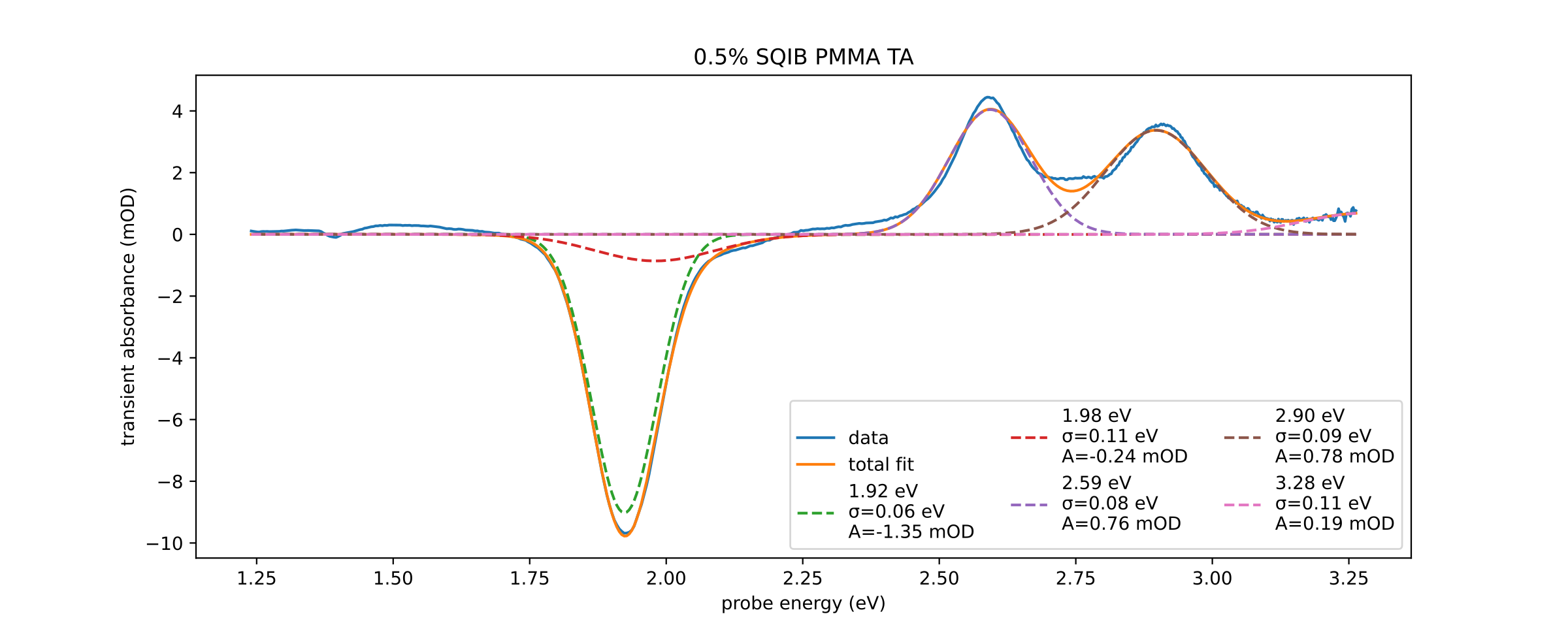}
    \caption{TA spectrum of 0.5~\% SQIB in PMMA with excitation at 650~nm. Chirp correction was performed using a second-order polynomial fit of the rising edges of visible features, then the average of the first ps of transient absorbance was fitted using six normalized Gaussian pulses. First value is the peak position, $A$ area under the respective Gaussian, $\sigma$ standard deviation. Note that the ESA peaks are given in terms of relative energy, to get the total energy the pump energy of 1.9~eV has to be added.}
    \label{fig:S7}
\end{figure}
\clearpage



\providecommand{\latin}[1]{#1}
\makeatletter
\providecommand{\doi}
  {\begingroup\let\do\@makeother\dospecials
  \catcode`\{=1 \catcode`\}=2 \doi@aux}
\providecommand{\doi@aux}[1]{\endgroup\texttt{#1}}
\makeatother
\providecommand*\mcitethebibliography{\thebibliography}
\csname @ifundefined\endcsname{endmcitethebibliography}
  {\let\endmcitethebibliography\endthebibliography}{}

\end{document}